\documentclass[aps,prb,twocolumn,superscriptaddress,floatfix]{revtex4-2}
\usepackage[utf8]{inputenc}
\usepackage{natbib}
\usepackage[colorlinks=true,linkcolor=blue,citecolor=blue,urlcolor=blue]{hyperref}
\usepackage{graphicx}
\usepackage{latexsym}
\usepackage{amssymb}
\usepackage{amsmath}
\usepackage{amsfonts}
\usepackage{amsthm}
\usepackage{bm}
\usepackage{floatrow}
\usepackage[caption=false]{subfig}
\usepackage{bbm}
\usepackage{enumitem}
\usepackage{hyperref}
\usepackage{lipsum}
\usepackage{braket}
\usepackage{tikz}
\usepackage[margin=1in]{geometry}
\usetikzlibrary{shapes.geometric, arrows.meta}
\usepackage{tikzit}
\usepackage{pgfplots}
\usepackage{comment}
\usepackage{diagbox}
\usepackage{ifthen}
\usepackage{ragged2e}
\usepackage{multirow}
\usepackage{mathtools}
\usepackage[export]{adjustbox}
\usetikzlibrary{arrows,decorations.pathreplacing,decorations.markings,arrows.meta,patterns,3d}
\usepackage[normalem]{ulem} 
\usepackage{cancel}
\usepackage{microtype}
\usepackage{booktabs}

\theoremstyle{definition}

\begin{document}

\title{Ginzburg-Landau Theory for Non-Invertible Symmetry-Breaking Transitions}
\date{\today}

\author{Vibhu Ravindran}
\affiliation{Department of Physics and Institute for Quantum Information and Matter, \mbox{California Institute of Technology, Pasadena, CA, 91125, USA}}
\author{Luisa Eck}
\affiliation{Department of Physics and Institute for Quantum Information and Matter, \mbox{California Institute of Technology, Pasadena, CA, 91125, USA}}
\author{Xie Chen}
\affiliation{Department of Physics and Institute for Quantum Information and Matter, \mbox{California Institute of Technology, Pasadena, CA, 91125, USA}}

\begin{abstract}
Recently, a generalized Landau paradigm has been proposed, in which a broad class of “unconventional” phase transitions lying beyond the original symmetry-breaking framework of Landau are understood as the symmetry-breaking transitions of \textit{generalized} symmetries. In particular, in (1+1)D, such “topological” or “deconfined” transitions can be mapped to the symmetry-breaking transitions of non-invertible symmetries. How to extract the universal dynamical properties of the transitions from this understanding?  In this paper, we follow the Ginzburg-Landau philosophy and develop a field-theoretic description of such critical points in terms of the fluctuations of local order parameters, now of non-invertible symmetries. The Symmetry Topological Field Theory formalism plays a central role in our analysis, allowing us to identify the algebra of local order parameters from the Lagrangian algebra of a (2+1)D topological order. We illustrate the key ideas and the full steps of this generalized Ginzburg-Landau procedure using the transitions between gapped phases with a simple non-invertible symmetry -- the $\mathrm{Rep}(S_3)$ symmetry.
\end{abstract}

\maketitle


\section{Introduction}


In the Landau paradigm of phases and phase transitions\cite{Landau1937}, different phases result from the different ways the symmetry of the system is broken. Phase transitions are driven by the fluctuation of an order parameter of the symmetry. As the order parameter develops a long range correlation and attains a non-zero expectation value, the symmetry is spontaneously broken and the system enters a symmetry-broken phase. This process is captured by the Ginzburg-Landau field theory which, in many cases, successfully predicts the universality class of the symmetry-breaking critical point.


While it is well-understood that, in quantum many-body systems, many phases and phase transitions lie beyond the Landau paradigm, recent development has led to the proposal of a generalized Landau paradigm\cite{Gaiotto2015,Hofman2019,Delacretaz2020,Iqbal2020,Moradi2023,Bhardwaj2024,McGreevy2023,Chen2025} which seeks to bring such unconventional phases and phase transitions back into the symmetry breaking framework, with generalized symmetries\cite{Gaiotto2015,McGreevy2023,Shao2024TASI,SCHAFERNAMEKI2024}. In particular, in $1+1$D, almost all of the transitions between gapped phases with discrete symmetries can be mapped to symmetry-breaking transitions as long as the symmetry is allowed to be non-invertible, i.e. the symmetry operators do not form a group. In Ref.~\cite{Chen2025}, we established such a mapping using the `generalized gauging' procedure such that the resulting transition is always from a partial symmetry breaking phase to the fully symmetry breaking phase of the (possibly) non-invertible symmetry .

Knowing that the transition is induced by the spontaneous breaking of a non-invertible symmetry does not immediately tell us about the universal properties of the transition. In this paper, we show that Landau's philosophy still applies. That is, the universal dynamical aspects of the transition are still determined by the fluctuation of \textit{local} order parameters, now of non-invertible symmetry. In Ref.~\cite{chen2026spontaneous}, we showed that, similar to the symmetry-broken phases of invertible symmetries, the symmetry-broken phases of non-invertible symmetries are characterized by the long-range correlation of \textit{local} order parameters. In this paper, we show that we can follow the Ginzburg-Landau procedure to write down a field theory to describe the symmetry-breaking transition of non-invertible symmetries. The degrees of freedom in the field theory are the order parameter fields that drive the transition and the allowed Lagrangian terms are combinations of such fields that are invariant under the non-invertible symmetry. Using the example of phase transitions in a system with the $\mathrm{Rep}(S_3)$ symmetry, we show that the field theory reproduces the expected critical points.

The fact that Landau's idea based on order parameters can be generalized to non-invertible symmetries may not seem obvious at first sight. After all, order parameters of a non-invertible symmetry are very different from those of invertible symmetries\cite{Bhardwaj2024}. For example, under a non-invertible symmetry action, local order parameters can transform into non-local operators, which is not possible with ordinary invertible symmetries. As a result, it is not entirely clear that the symmetry-breaking transitions can be described by a field theory with local fields alone. One essential tool that helps us to understand how this can be done is the Symmetry Topological Field Theory (SymTFT) formalism\cite{Kong2015,Ji2020,Kong2020,Bhardwaj2020,Pulmann2021,Gaiotto2021,Lichtman2021,kong2020mathematical,Kong2020classification,Apruzzi2023,Chatterjee2023symmetry,Moradi2022topological,Freed2023topological,Lin2023,Kong2018,Kong2021,Kong2022one,Kong2022categories,Kong2024categories,Xu2024,Bhardwaj2025}. SymTFT relates the algebra of the order parameters to the Lagrangian algebra of a $2+1$D topological order, which is a crucial input to the field theory we develop in this paper. 

The paper is organized as follows. In section~\ref{sec:invertible}, we start by reviewing the basic formalism of Ginzburg-Landau for invertible symmetries. We use the nonabelian invertible symmetry $S_3$ as an example to illustrate how the original formulation of Ginzburg-Landau is reinterpreted in the SymTFT formalism, where the algebra of the order parameters can be derived from the algebra of anyons in the bulk of a $2+1$D topological order and the condensation at its gapped boundary. In section~\ref{sec:non-invertible}, we use the SymTFT formalism to tackle the symmetry-breaking transition of the $\mathrm{Rep}(S_3)$ non-invertible symmetry and show how the resulting field theory correctly captures the universality class at the critical points. Phases and phase transitions in systems with the $\mathrm{Rep}(S_3)$ symmetry have also been studied in Ref.~\cite{Choi2023,Chatterjee2024,Bhardwaj2025,Bhardwaj2024,Bhardwaj2026}. The key ideas and full steps of this generalized Ginzburg-Landau procedure are summarized in section~\ref{sec:discussion}, where we also discuss its limitations and related open directions. 




\section{Invertible Ginzburg-Landau}
\label{sec:invertible}

\subsection{Review of Ginzburg-Landau}
\label{sec:invertible_review}

In this section, we review the basic procedure to write down a Ginzburg-Landau field theory for describing the symmetry-breaking transition of an ordinary invertible symmetry. This procedure can be found in many textbooks (see for example\cite{Landau1980,Kardar2007}). Readers familiar with the Ginzburg-Landau formulation can skip this section and continue with section~\ref{sec:invertible_SymTFT}.

The essential element in formulating a Ginzburg-Landau field theory is the local order parameter driving the symmetry-breaking transition. For a system with symmetry of group $G$, its local order parameters transform as irreducible representations $\rho$ of the group. Each irrep $\rho$ of dimension $d$ transforms as a $d$-dimensional vector space under the symmetry. In the symmetric phase, the expectation value of (each component of) $\rho$ has to be zero. In the symmetry-breaking phase, the irrep can develop a nonzero expectation value along a particular direction, breaking the full symmetry down to a subgroup $H (\subset G)$.  

The basic assumption of Ginzburg-Landau is that, close to the transition, the order parameter varies smoothly with space and time and can be represented by continuous fields. The assignment of field variables to the order parameter depends on the type of the irrep. 
\begin{enumerate}[label=\alph*)]
\item If the irrep is real, i.e. $\rho \otimes \rho$ contains the trivial irrep and the singlet state is symmetric under the exchange of the two $\rho$'s, then there is a basis for the irrep such that the symmetry operators are represented by real matrices. The basis components in this basis can be assigned real fields $\phi_i(\vec{x},t)$, $i=1,...,d$. If rotated to other bases, the basis fields may need to be complex but they are related in a way that only $d$ real independent field variables are necessary. 
\item If the irrep is complex, i.e. $\rho \otimes \rho$ does not contain the trivial irrep, then each basis component of $\rho$ can be assigned a complex field $\psi_i(\vec{x},t)$, $i=1,...,d$. The number of independent field variables doubles. 
\item If the irrep is pseudo-real, i.e. $\rho \otimes \rho$ contains the trivial irrep but the singlet state is anti-symmetric under the exchange of the two $\rho$'s, each basis component can still be assigned a complex field $\psi_i(\vec{x},t)$, $i=1,...,d$ but there is a gauge redundancy among the components. 
\end{enumerate}
In section~\ref{sec:non-invertible}, we will discuss how this set of rules can be generalized to order parameters of non-invertible symmetries. The real basis in case a) can also be identified from the requirement that independent terms in the Lagrangian have to be real. We will use this point of view to identify the real basis for the order parameters of $S_3$ and we will generalize this point of view to apply to order parameters of non-invertible symmetries in section~\ref{sec:non-invertible}. 

Using the continuous fields assigned to the order parameters, one can form Lagrangian terms that are invariant under the global symmetry action. The Ginzburg-Landau field theory is composed of such terms, usually truncated to terms of the lowest few orders. Varying the coefficient of a relevant term in the field theory takes the system across the transition and the field theory in principle contains all the universal information about the transition, including the critical properties. In practice, the Ginzburg-Landau field theory is most useful in high dimensions (e.g. $>3+1$D) when the mean-field approximation gives accurate results. In lower dimensions, like $1+1$D, the field theory is usually strongly coupled and has to be solved with special tools (e.g. numerical simulation or mapping to free fermion theories). 

Let's illustrate the process using the example of the $S_3$ symmetry-breaking transitions. This is to be compared and contrasted with the procedure for the non-invertible $\mathrm{Rep}(S_3)$ symmetry discussed in section~\ref{sec:non-invertible}. $S_3$ is the nonabelian symmetry group of a triangle. It is generated by a three-fold rotation operation $f$ ($f^3=1$) and a reflection operation $d_1$ ($d_1^2=1$). $f$ and $d_1$ satisfy the relation $(fd_1)^2 = 1$. We will use the notation $\bar{f}=f^2$. $S_3$ has an order-three subgroup generated by $f$ and three order-two subgroups generated by $d_1$, $d_2=fd_1$, and $d_3=\bar{f}d_1$. Accordingly, there are four gapped phases in systems with $S_3$ symmetry: the $S_3$ symmetric phase, the $S_3$ to $Z_3$ symmetry-breaking phase, the $S_3$ to $Z_2$ symmetry-breaking phase, and the $S_3$ to trivial $Z_1$ symmetry-breaking phase.  

$S_3$ has three irreducible representations: the identity irrep $A$, the sign irrep $B$, and the two-dimensional irrep $C$, which satisfy the fusion rule of \footnote{We use the notation $A$, $B$, $C$, $f$, and $d_1$ to be consistent with the Quantum Double\cite{Kitaev2003} notation.}
\begin{equation}
B \times B = A, \ B \times C = C, \ C \times C = A + B + C
\label{eq:ABCfusion}
\end{equation}
The fluctuation of either $B$ or $C$ type order parameter drives the transition from a phase with a larger symmetry to one with a smaller symmetry. 

Let's consider first the case of $B$. $B$ is a one-dimensional real irrep. Therefore, it can be assigned a real field $\phi_b$, which transforms as
\[
\phi_b \xrightarrow{\text{$f$}} \phi_b, \  \phi_b \xrightarrow{\text{$d_1$,$d_2$,$d_3$}} -\phi_b
\]
The Ginzburg-Landau field theory involving $B$ hence contains all the even power terms of $\phi_b$
\begin{equation}
\mathcal{L}_B = K_{\tau} (\partial_{\tau}\phi_b)^2 + K_x (\partial_x \phi_b)^2 + a \phi_b^2 + b \phi_b^4 + ...
\label{eq:LB}
\end{equation}
where $...$ includes higher order terms and $K_{\tau}$, $K_x$, $a$, $b$, ... are real coefficients. If a $B$ type order parameter acquires a nonzero expectation value, the $d$ symmetries are broken. Therefore, $\mathcal{L}_B$ describes the (Ising) transition from the $S_3$ symmetric phase to the $S_3$ to $Z_3$ symmetry-breaking phase or the $S_3$ to $Z_2$ symmetry-breaking phase to the $S_3$ to $Z_1$ symmetry-breaking phase. It is known that in both cases, the critical point is described by the 2D Ising conformal field theory.

$C$ is also a real irrep, but of two dimensions. When the $S_3$ group is interpreted as the transformation of the two-dimensional plane (where the triangle is embedded), the transformation matrices are real in the $x$ and $y$ basis. Therefore, one can assign two real continuous fields $\phi_{c_x}$ and $\phi_{c_y}$ to describe the fluctuation of a $C$ type order parameter. Their complex linear combinations
\[
\psi_{c} = \phi_{c_x} + i\phi_{c_y}, \ \psi_{\bar{c}} = \phi_{c_x} - i\phi_{c_y}
\]
are complex conjugates of each other and transform under the $S_3$ symmetry as
\[
\begin{array}{l}
\psi_{c} \xrightarrow{f}  \omega \psi_{c}, \ \psi_{\bar{c}} \xrightarrow{f} \bar{\omega} \psi_{\bar{c}}, \\ \psi_{c} \xrightarrow{d_1} \psi_{\bar{c}}, \ \psi_{c} \xrightarrow{d_2} \bar{\omega}\psi_{\bar{c}}, \ \psi_{c} \xrightarrow{d_3} \omega \psi_{\bar{c}}
\end{array}
\]
where $\omega = e^{i 2\pi/3}$.

The assignment of real / complex fields to the order parameter components can also be determined from the requirement that independent Lagrangian terms are real. At second order, the Lagrangian term takes the form $\phi_{c_x}^2 + \phi_{c_y}^2$. For the term to be always real, $\phi_{c_x}$ and $\phi_{c_y}$ both have to be real. Equivalently, the second order term can be written as $\psi_c\psi_{\bar{c}}$. For this term to be real, $\psi_c$ can be complex but $\psi_{\bar{c}}$ should be its complex conjugate, which leads to the same field assignment. This point of view is redundant for invertible symmetry, but will be useful when we assign fields to the order parameters of non-invertible symmetries. 

Denote $\bar{\psi}_c = \psi_{\bar{c}} = \psi^*_{c}$. $\phi_{c_x}^2+\phi_{c_y}^2 =  |\psi_c|^2$ is the only (non-derivative) second order term invariant under the full symmetry. At third order, we have the term $\psi_c^3 + {\bar{\psi}_c}^3$ and at fourth order, we have $|\psi_c|^4$. Therefore, at lowest order, the Ginzburg-Landau theory due to the fluctuation of an order parameter of the $C$ type looks like
\begin{equation}
\begin{array}{lll}
\mathcal{L}_C & = &  K_{\tau} |\partial_{\tau}\psi_c|^2 + K_x |\partial_x \psi_c|^2 + \\
& & a |\psi_c|^2 + b (\psi_c^3 + \bar{\psi}_c^3) + c |\psi_c|^4 + ...
\end{array}
\label{eq:LC}
\end{equation}
With $b<0$, a mean-field analysis shows that, when $a$ is sufficiently small, a classical minimum of $\mathcal{L}_C$ is achieved at a $\psi_c$ that breaks the rotational symmetry generated by $f$ but preserves one of the reflection symmetries $d_i$. Therefore, $\mathcal{L}_C$ describes the symmetry-breaking transition from $S_3$ to $Z_2$. It is well known that the critical point is described by the 2D 3-state Potts conformal field theory.

With the $C$ type order parameter, there is another interesting case where the $S_3$ symmetry is already spontaneously broken down to $Z_3$ before $C$ starts to fluctuate. That is, $\phi_b$ already attains a nonzero expectation value, which leads to an extra spatial derivative term in the Lagrangian
\begin{equation}
\mathcal{L}_{C,B} = \mathcal{L}_C + b_1 \phi_b \left(\psi_c\partial_x\bar{\psi}_c-\bar{\psi}_c\partial_x\psi_c \right) + ...
\label{eq:LCB}
\end{equation}
$\phi_b \left(\psi_c\partial_x\bar{\psi}_c-\bar{\psi}_c\partial_x\psi_c \right)$ is called the chiral term because it breaks the reflection symmetry of $\psi_c \to \bar{\psi}_c$. ``..." indicates higher order chiral terms like $\phi_b\left(\psi_c^3-\bar{\psi}_c^3\right)$. It is known that $\mathcal{L}_{C,B}$ describes the transition from the $S_3$ to $Z_3$ symmetry-breaking phase to the fully symmetry-breaking ($S_3$ to $Z_1$) phase with a non-conformal chiral critical point (with weak chirality)\cite{Huse1982,Chepiga2019,Maceira2022}.

\subsection{SymTFT re-interpretation}
\label{sec:invertible_SymTFT}

In this section, we review how a $1+1$D system with invertible group $G$ symmetry can be recast in a `sandwich' structure using the Symmetry Topological Field Theory (SymTFT) formalism. With this setup, we will discuss in section~\ref{sec:invertible_Op}, how the algebra of the order parameter and the invariant terms in the Lagrangian can be extracted from the categorical data describing the Quantum Double\cite{Kitaev2003} type topological order and its `rough' boundary used in the sandwich construction. This understanding will be generalized to all other types of $2+1$D topological orders and their gapped boundary conditions in Section~\ref{sec:non-invertible}, which leads to the formulation of the generalized Ginzburg-Landau field theory for non-invertible symmetries. 

\begin{figure}[th]
\begin{center}
\includegraphics[width= 0.8\textwidth]{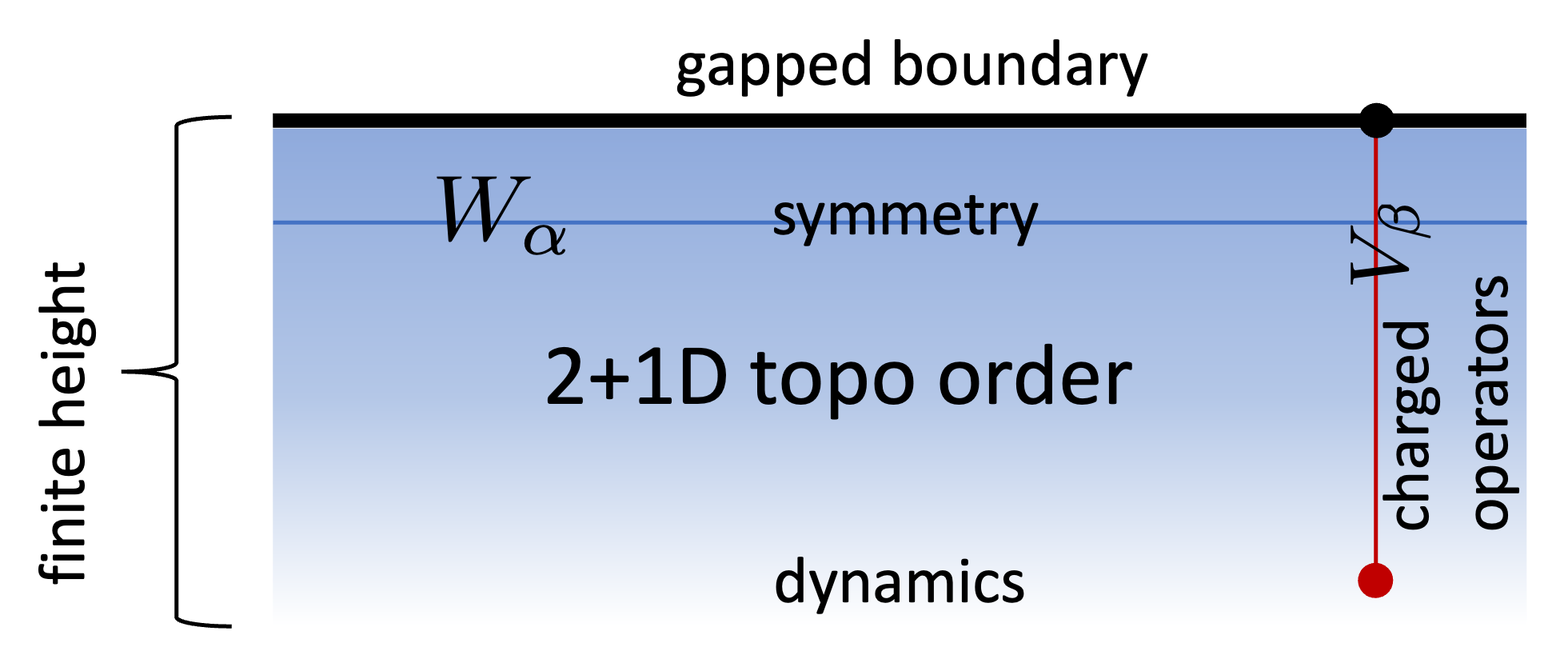}
\caption{In the Symmetry Topological Field Theory, a $1+1$D system is realized in a sandwich structure with a $2+1$D topological order in the bulk. The top boundary is gapped, fixed, and determines the symmetry of the system. The bottom boundary contains all the dynamics of the system and is at a finite distance from the top boundary. The horizontal string operators $W_{\alpha}$ become the symmetry operators of the $1+1$D system. The vertical string operators $V_{\beta}$ become local charged operators if $\beta$ is condensed on the top boundary.} 
\label{fig:symTFT}
\end{center}
\end{figure}

In the SymTFT formalism, a $1+1$D system is realized as a sandwich structure with a $2+1$D topological bulk, as shown in Fig.~\ref{fig:symTFT}. The top boundary is set to be in a gapped state through the condensation of certain bulk anyons. The bottom boundary is left open to host the dynamics of the system. With a finite distance between the top and bottom boundary, the sandwich effectively represents a $1+1$D system. We consider only $1+1$D systems with a periodic boundary condition, hence the left and right end of the sandwich are identified. The advantage of having the sandwich structure is that the dynamics at the bottom boundary is spatially separated from the action of the symmetry. The string operators that run parallel to the top and bottom boundaries ($W_{\alpha}$ associated with the anyon $\alpha$) become the ($0$-form) symmetry of the $1+1$D system. The ones that correspond to anyons condensed on the top boundary take fixed values while the others become true symmetries of the $1+1$D system that act non-trivially. The vertical string operator $V_{\beta}$ which tunnels a condensed anyon $\beta$ out of the top boundary can have nontrivial commutation relations with the symmetry operators $W_{\alpha}$. Moreover, due to the finite height of the sandwich structure, the vertical strings are effectively local operators in the $1+1$D system. Therefore, $V_{\beta}$ corresponds to local charged operators under the symmetry. There is one $\beta$ anyon at each end of the $V_{\beta}$ string (dots in Fig.~\ref{fig:symTFT}). The top one (the black dot) merges into the condensate at the top boundary while the bottom one (the red dot) transforms under the $W_{\alpha}$ symmetry operators and participates in the dynamics near the bottom boundary.  

To represent a $1+1$D system with a $0$-form symmetry of (finite) group $G$, the $2+1$D bulk topological order can be chosen to be the Quantum Double of $G$\cite{Kitaev2003}. The anyons in the bulk include gauge charges labeled by the irreps of $G$ and gauge fluxes labeled by the conjugacy classes of $G$. The top boundary is set to be the `rough' boundary where all the gauge charges condense. The horizontal string operator $W_{\alpha}$ of gauge charges acts trivially due to the condensation while the horizontal string operators of gauge fluxes act nontrivially and form the $G$ symmetry of the sandwich. The vertical string operators of the gauge charges become the local charged operators -- the local order parameters -- of the symmetry. 

In the case of $G=S_3$, the bulk of the SymTFT sandwich has the $S_3$ Quantum Double topological order. Following the notation in Ref.~\cite{Kitaev2003,Beigi2011}, the anyons in the bulk include the gauge charges $A$, $B$, and $C$ labeled by the irreps of $S_3$, the gauge fluxes $D$ and $F$ labeled by the reflection and the rotation conjugacy classes of $S_3$, and the dyons $E$ and $G$. The anyons $A$ to $G$ form a braided fusion category. Some of the basic data of this category can be found in Appendix~\ref{app:S3}. Here we highlight some features. The $A$, $B$ anyons have quantum dimension 1 while $C$ has quantum dimension 2, corresponding to the internal dimensions of the irreps. $A$, $B$ and $C$ are closed under fusion, as shown in Eq.~\ref{eq:ABCfusion}. The D and F anyons are three and two-dimensional respectively, corresponding to the three group elements ($d_1$, $d_2$, $d_3$) in the reflection conjugacy class and the two group elements ($f$ and $\bar{f}$) in the rotation conjugacy class. 

When the gauge charges $A$, $B$, and $C$ are condensed on the top boundary, the top boundary is gapped. The gapped boundary is described by a `Lagrangian sub-algebra'\footnote{The word `Lagrangian' here is not related to the Lagrangian in the field theory.} of the bulk anyon fusion category
\[
\mathcal{A}_1 = A + B + 2C
\]
The summands in $\mathcal{A}_1$ ($A$, $B$, $C$) give the types of anyons that are condensed on the boundary, and the coefficient in front of each term indicates the number of internal dimensions of each anyon that are condensed. On the `rough' boundary of a Quantum Double topological order, where all gauge charges condense, the coefficient is always equal to the full dimension of each irrep, indicating that all dimensions of the corresponding anyons are condensed. In later discussions, we will see that this is not always true on gapped boundaries of general $2+1$D topological states. That is, when an anyon condenses on a gapped boundary, it is possible that only some of its internal dimensions condense. This will have important consequences for the algebra of the order parameters of non-invertible symmetries and the resulting Ginzburg-Landau field theory, as we explore in Section~\ref{sec:non-invertible}.

Due to the condensation, the horizontal string operator $W_{D}$ splits into three components $W_{d_1}$, $W_{d_2}$ and $W_{d_3}$. $W_{F}$ splits into two components $W_{f}$ and $W_{\bar{f}}$. Together with $W_A$, they form the $S_3$ symmetry of the sandwich. The vertical string operators $V_B$ and $V_C$ transform nontrivially under the symmetry operators and form the local order parameters of the symmetry. In particular, 
\[
W_{f}V_BW_{f}^{\dagger} = V_B, \ W_{d_i}V_BW_{d_i}^{\dagger} = -V_B
\]
Unsurprisingly, $V_B$ transforms as the $B$ irrep of $S_3$. The string operator $V_C$ is slightly more complicated. It carries one $C$ anyon at each end, each with a two-dimensional internal Hilbert space. If we label the two dimensions of the top $C$ anyon as $\tilde{C}_k$ and that of bottom $C$ anyon as $C_i$, $k,i=x,y$, the string operators transform as
\[
W_g V^{\tilde{c}_k}_{c_i} W_g^{\dagger} = \sum_j T^{(C)}_{ij}(g)V^{\tilde{c}_k}_{c_j}
\]
where $T^{(C)}_{ij}(g)$ is the two-dimensional matrix representing the group element $g$ in the $C$ irrep. The top index $\tilde{C}_k$ does not transform under the symmetry, while the bottom index transforms as the $C$ irrep of $S_3$. In this example, the top index simply labels different copies of the same type of order parameter. In Section~\ref{sec:non-invertible}, we will see that, for non-invertible symmetries, the top index plays a very important role of constraining the algebra of the order parameters. It will also become clear that for invertible symmetries, the top index can be safely ignored when considering the algebra of the order parameters. 

\subsection{Order Parameter Algebra from SymTFT}
\label{sec:invertible_Op}

Using the SymTFT setup described above, we show how to extract the operator algebra (the fusion rules) of the order parameters and the invariant terms in the Ginzburg-Landau Lagrangian from the categorical data describing the Quantum Double type topological order and its gauge charge condensed boundary. 

\begin{figure}[th]
\begin{center}
\includegraphics[width= 1.0\textwidth]{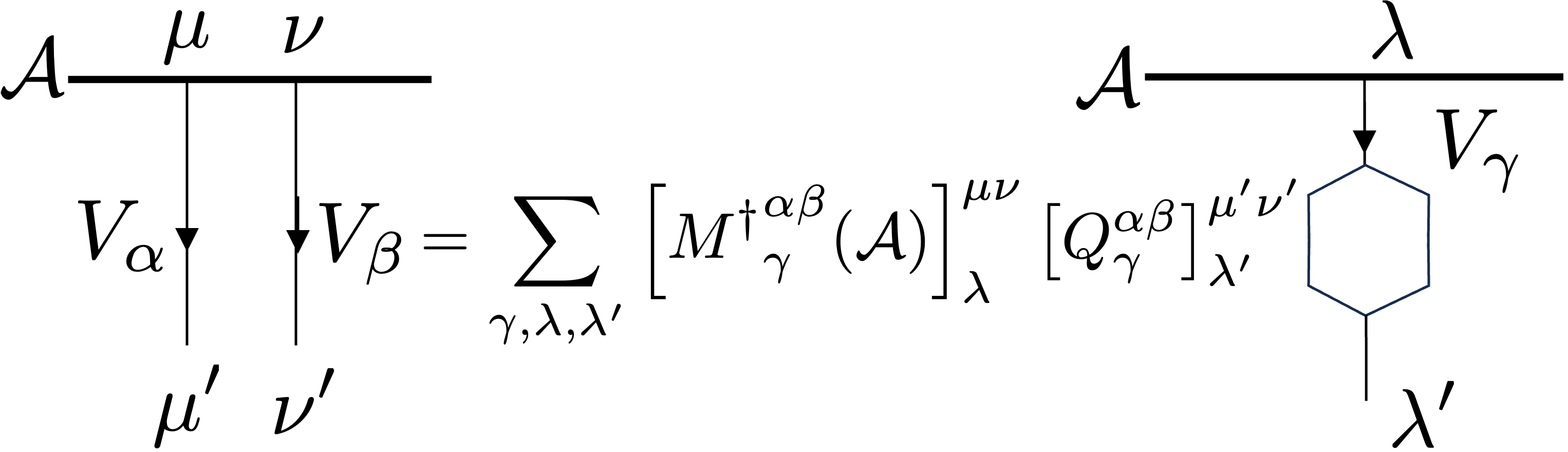}
\caption{The multiplication (fusion) of two order parameter operators in a SymTFT sandwich. $\mathcal{A}$ labels the Lagrangian algebra describing the top boundary of the sandwich (thick black line). $V_{\alpha}$, $V_{\beta}$ and $V_{\gamma}$ are order parameter operators labeled by anyons condensed in $\mathcal{A}$. $M$ is the multiplication coefficient in $\mathcal{A}$. $Q$ is the $3j$ symbol of the fusion of anyons in the bulk.} 
\label{fig:OpAl}
\end{center}
\end{figure}

As explained in section~\ref{sec:invertible_SymTFT}, the vertical string operators $V_{\alpha}$ which tunnel condensed anyons from the top boundary of the SymTFT sandwich become local order parameters of the system. If we have two such order parameters, their multiplication (fusion) is given in Fig.~\ref{fig:OpAl}. $\mathcal{A}$ is the Lagrangian algebra describing the gapped top boundary (thick black line). $\mu$, $\nu$, $\lambda$ label the condensed dimensions of the $\alpha$, $\beta$ and $\gamma$ anyons at the top boundary. $\mu'$, $\nu'$, $\lambda'$ label the full internal dimensions of the $\alpha$, $\beta$, $\gamma$ anyons in the topological bulk. (In the following discussion, we assume that the anyons associated with the order parameters have integer quantum dimensions. In section~\ref{sec:discussion}, we discuss how the procedure illustrated in this paper might be generalized to apply to general SymTFT systems where the quantum dimension of the condensed anyons can be non-integer.) $Q$ is the $3j$ symbol of the fusion of $\alpha$ and $\beta$ into $\gamma$ in the topological bulk. $M$ is the multiplication coefficient of $\alpha$ and $\beta$ into $\gamma$ on the top boundary. Once we know which internal dimensions of $\alpha$, $\beta$ and $\gamma$ are condensed, $M$ can be obtained from $Q$ by projection onto those dimensions. This operator algebra described here is similar to that discussed in section 5 (Eq. 5.2) of Ref.~\cite{cong2016topological}. The difference is that here we leave the bottom indices open and do not project them onto a fixed bottom boundary.

In the case of a sandwich structure with the Quantum Double bulk topological order and the gauge charge condensed top boundary, $Q$ is simply the Clebsch-Gordan coefficient of the irreps of the group, and $M$ is the same as $Q$ since all the internal dimensions of the gauge charges are condensed. When $G=S_3$, denote the one dimension of irrep $A$ as $\ket{a}$, the one dimension of irrep $B$ as $\ket{b}$ and the two dimensions of irrep $C$ as $\ket{c}$ and $\ket{\bar{c}}$, the Clebsch-Gordan coefficients we use in this paper are given by
\[
\begin{array}{l}
|bb\rangle \to |a\rangle, \ |bc\rangle \to |c\rangle, \ |b\bar{c}\rangle \to -|\bar{c}\rangle \\
\frac{1}{\sqrt{2}}\left(|c\bar{c}\rangle + |\bar{c}c\rangle\right) \to |a\rangle, \ \frac{1}{\sqrt{2}}\left(|c\bar{c}\rangle - |\bar{c}c\rangle\right) \to |b\rangle \\
|cc\rangle \to |\bar{c}\rangle, \ |\bar{c}\bar{c}\rangle \to |c\rangle
\end{array}
\]

To write down a Ginzburg-Landau field theory, we need to assign continuous fields to the order parameters. We assign fields components to each dimension of the bottom indices $\mu'$, $\nu'$ etc of the $V$ operators, since they are the ones that transform under the symmetry and generate dynamics of the phase transition. Based on the same argument used above, we associate one real field $\phi_b$ with $V_B$ (hence $\ket{b}$) and two real fields $\phi_{c_x}$, $\phi_{c_y}$ with the components $\ket{c_x}$ and $\ket{c_y}$ of $V_C$, which can be combined into a single complex field $\psi_{c} = \phi_{c_x} + i \phi_{c_y}$ corresponding to the $\ket{c}$ component of $V_C$. The $\ket{\bar{c}}$ component is associated with the complex conjugate $\bar{\psi}_c$ field.

\begin{figure}[th]
\begin{center}
\includegraphics[width= 0.8\textwidth]{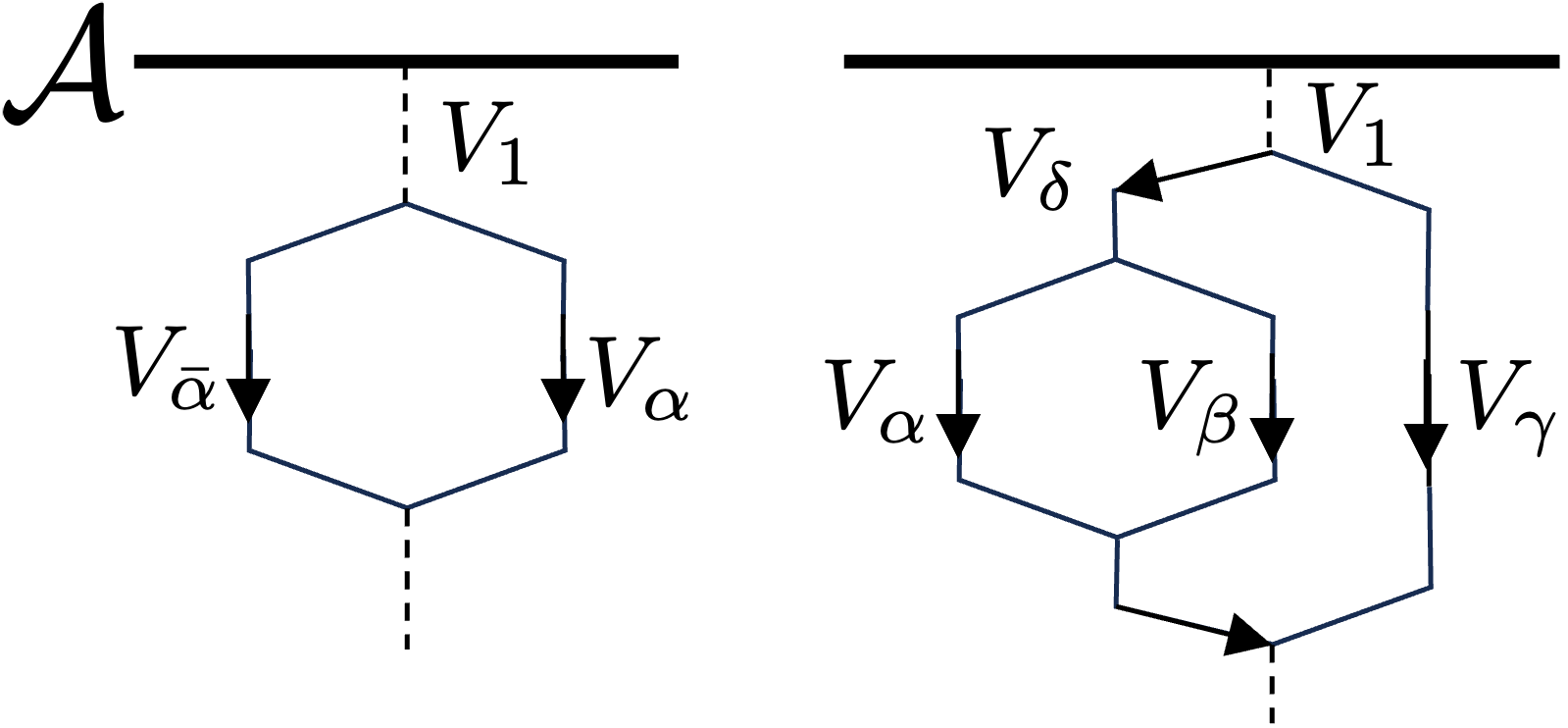}
\caption{(a) Order two and (b) order three fusion trees of the order parameter $V$ operators. that fuse into the trivial anyon $1$.} 
\label{fig:FusionTree}
\end{center}
\end{figure}

The terms in the Ginzburg-Landau Lagrangian are combinations of local order parameters that are invariant under the symmetry action. In the SymTFT setup, such terms correspond to fusion trees of the $V$ operators that fuse into the trivial anyon, as shown in Fig.~\ref{fig:FusionTree}. With fixed fusion branches ($\alpha$, $\beta$, $\gamma$, etc.) and fixed fusion multiplicity (if the multiplicity is nontrivial), the top and the bottom part of the fusion tree decouples. The bottom part gives combinations of the dynamical fields (with coefficient $Q$) that are invariant under the symmetry. The top part gives simply a multiplicative factor. In the case of $S_3$ (or any invertible symmetry), the top part does not play any important role and can be ignored when putting the terms together into a Ginzburg-Landau Lagrangian. For non-invertible symmetries, the top part of the fusion tree can constrain possible terms in the Lagrangian because the $M$ coefficient can be zero when the $Q$ coefficient is not. We will illustrate how this happens in section~\ref{sec:non-invertible_Op} using the $\mathrm{Rep}(S_3)$ example. 

Let's list all the invariant terms (up to the fourth order) from the combination of the $B$ and $C$ type order parameters, first as quantum states of operators (linear combinations of the $\mu'$, $\nu'$ indices given by the fusion tree), then as terms in the Lagrangian. 

At second order, there are two fusion trees and correspondingly two Lagrangian terms. 

\begin{figure}[h]
\begin{center}
\includegraphics[width= 0.4\textwidth]{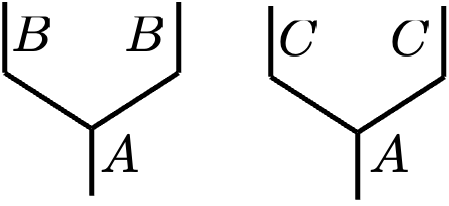}
\end{center}
\end{figure}

If we write the single component of $B$ as $\ket{b}$ and the two components of $C$ as $\ket{c}$ and $\ket{\bar{c}}$, the two fusion trees correspond to quantum states
\[
\ket{bb}, \ \ket{c\bar{c}}+\ket{\bar{c}c}
\]
Replacing the operators with their assigned fields, we obtain the quadratic Lagrangian terms
\[
\phi_b^2, \ |\psi_c|^2
\]
If we allow for spatial and time derivatives (assuming that the $S_3$ symmetry is an internal, not a spatial, symmetry), we obtain terms like
\[
\left(\partial_{\tau} \phi_b\right)^2, \left(\partial_x \phi_b\right)^2, |\partial_{\tau} \psi_c|^2, |\partial_x \psi_c|^2
\]

At third order, there are two fusion trees, one involving three $C$'s and the other involving two $C$'s and one $B$. 

\begin{figure}[h]
\begin{center}
\includegraphics[width= 0.5\textwidth]{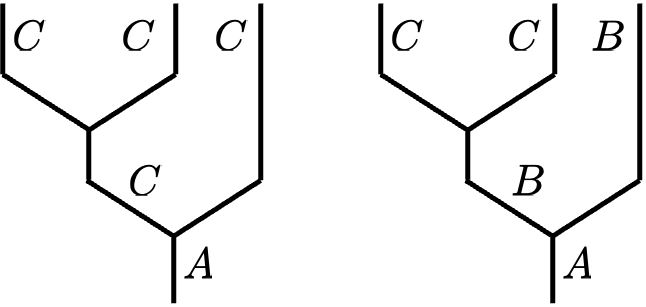}
\end{center}
\end{figure}

The two fusion trees correspond to quantum states
\[
\ket{ccc}+\ket{\bar{c}\bar{c}\bar{c}}, \ \left(\ket{c\bar{c}}-\ket{\bar{c}c}\right)\ket{b}
\]
From the first fusion tree, we get the third order Lagrangian term
\[
\psi_c^3 + \bar{\psi}_c^3
\]
The second fusion tree would be reduced to zero if we replace the two $C$'s with uniform fields. However, if we associate each with a different field, e.g. by allowing spatial variation from one $C$ to the other, we obtain the term
\[
\phi_b \left(\psi_c\partial_x\bar{\psi}_c-\bar{\psi}_c\partial_x\psi_c\right)
\]
which plays a crucial role in the chiral $Z_3$ phase transition\cite{Huse1982,Chepiga2019,Maceira2022}.

At fourth order, there are six fusion trees, as shown below. 

\begin{figure}[h]
\begin{center}
\includegraphics[width= 0.8\textwidth]{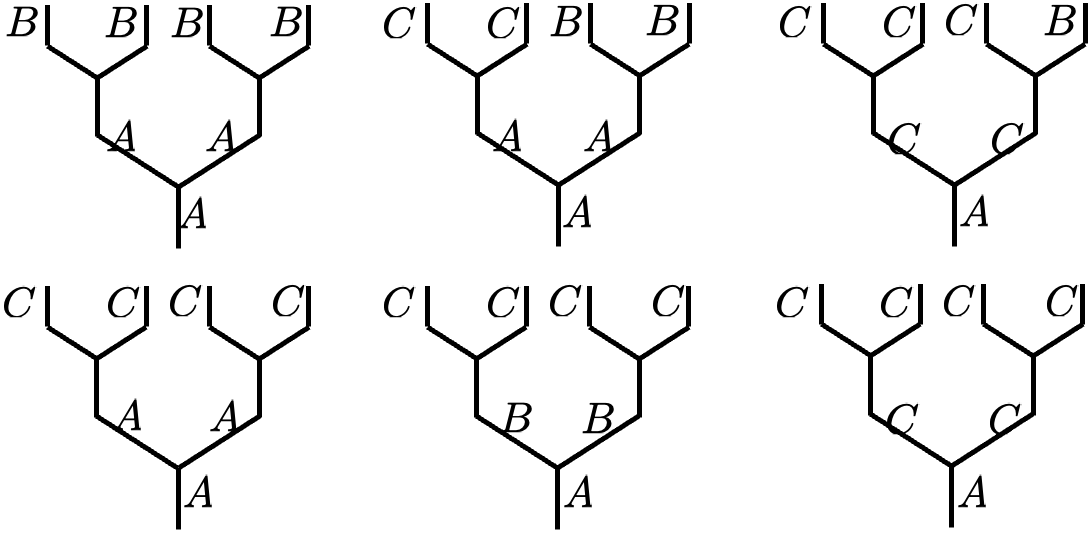}
\end{center}
\end{figure}

The fusion trees correspond to quantum states
\[
\begin{array}{l}
\ket{bbbb}, \left(\ket{c\bar{c}}+ \ket{\bar{c}c}\right)\ket{bb}, \\ 
\left(\ket{ccc}-\ket{\bar{c}\bar{c}\bar{c}}\right)\ket{b}, \left(\ket{c\bar{c}}+ \ket{\bar{c}c}\right)\left(\ket{c\bar{c}}+ \ket{\bar{c}c}\right), \\
\left(\ket{c\bar{c}}- \ket{\bar{c}c}\right)\left(\ket{c\bar{c}} - \ket{\bar{c}c}\right), \ket{cc\bar{c}\bar{c}} + \ket{\bar{c}\bar{c}cc}
\end{array}
\]
The fusion trees translate into Lagrangian terms
\[
\begin{array}{lll}
\phi_b^4, & |\psi_c|^2\phi_b^2, & \left(\psi_c^3-\bar{\psi}_c^3\right)\phi_b,\\ |\psi_c|^4, & \left(\psi_c\partial_x\bar{\psi}_c - \bar{\psi}_c\partial_x\psi_c\right)^2, & |\psi_c|^4
\end{array}
\]
Combining the terms listed above, we can obtain the Ginzburg-Landau description of the transitions. 

\subsection{Transitions in SymTFT}
\label{sec:invertible_transition}

In the SymTFT sandwich structure, where the bulk topological order and the gapped top boundary determine the symmetry of the system, the dynamics of the system is determined by what happens at the bottom boundary. When the bottom boundary is also gapped, the system is in a gapped phase. When the bottom boundary changes from one gapped boundary condition to a different gapped boundary condition, the system transitions from one gapped phase to another gapped phase.

For the Quantum Double type topological order of group $S_3$, there are four types of gapped boundaries, described by the Lagrangian algebras
\begin{equation}
\begin{array}{l}
\mathcal{A}_1 = A + B + 2C \\
\mathcal{A}_2 = A + B + 2F \\
\mathcal{A}_3 = A + C + D \\
\mathcal{A}_4 = A + D + F
\end{array}
\label{eq:A1-A4}
\end{equation}
When used as the bottom boundary (with the top boundary being $\mathcal{A}_1$), they correspond respectively to the four gapped phases of $S_3$. With $\mathcal{A}_1$, both $B$ and $C$ type order parameters are condensed (take on nonzero expectation value). Therefore, the $S_3$ symmetry is fully broken. With $\mathcal{A}_2$, only the $B$ type order parameter is condensed; therefore the $S_3$ symmetry is broken down to $Z_3$. With $\mathcal{A}_3$, only the $C$ type order parameter is condensed; therefore the $S_3$ symmetry is broken down to $Z_2$. With $\mathcal{A}_4$, neither of the order parameters is condensed. Therefore, the $S_3$ symmetry is not broken and $\mathcal{A}_4$ represents the $S_3$ symmetric phase. 

Let's look more carefully at how the anyons condense on the gapped boundary before discussing the phase transitions. While $C$ is condensed in both $\mathcal{A}_1$ and $\mathcal{A}_3$, it condenses in very different ways. In $\mathcal{A}_1$, both dimensions of $C$ are condensed. Therefore, in the fully symmetry broken phase, the expectation value of a $C$ type order parameter can point in any direction in the two-dimensional internal space. In $\mathcal{A}_3$, only one dimension of $C$ condenses. That is, a $C$ type order parameter can only point in specific directions. It cannot vary continuously in the whole internal space. This is consistent with the fact that in the $S_3$ to $Z_2$ symmetry breaking phase, since the reflection $Z_2$ symmetry is preserved, a $C$ type order parameter can only point in the $\ket{c}+\ket{\bar{c}}$, $\ket{c}+\omega\ket{\bar{c}}$, or $\ket{c}+\bar{\omega}\ket{\bar{c}}$ direction, $\omega = e^{i2\pi/3}$. The condensable dimension of the anyon labeling the order parameter is hence directly related to the direction the order parameter points to and takes nonzero expectation value in a symmetry-broken phase. In the following discussion, we will use these terms in an interchangeable way.

A similar analysis can be applied to the condensable dimensions of $D$ and $F$ and we will get to that in section~\ref{sec:non-invertible}. The understanding about condensable dimensions of the anyons on different gapped boundaries is used in Ref.~\cite{Li2026} to construct the lattice realization of the Quantum Double model with gapped boundaries and will be important for us to correctly formulate the field theories describing different transitions. 


Let's now consider the $S_3$ symmetry breaking transitions case by case. Consider first the $S_3$ symmetric to $S_3$ to $Z_3$ symmetry-breaking transition (bottom boundary $\mathcal{A}_4$ to $\mathcal{A}_2$). Comparing $\mathcal{A}_4$ and $\mathcal{A}_2$, we see that $B$ becomes condensed after the transition. Therefore, we expect the transition to be driven by the fluctuation of the order parameter $V_B$. The $C$ anyon is not condensed, either before or after the transition. Therefore, the expectation value of $V_C$ should be zero throughout. For all the Lagrangian terms listed in section~\ref{sec:invertible_Op}, we set $\psi_c =0$ and sum over the terms involving $\phi_b$. The resulting Lagrangian is
\[
\mathcal{L}_B = K_{\tau} (\partial_{\tau}\phi_b)^2 + K_x (\partial_x \phi_b)^2 + a \phi_b^2 + b \phi_b^4 + ...
\]
as expected. 

Consider next the $S_3$ symmetric to $S_3$ to $Z_2$ symmetry-breaking transition (bottom boundary $\mathcal{A}_4$ to $\mathcal{A}_3$). Comparing $\mathcal{A}_3$ and $\mathcal{A}_4$, we see that $C$ becomes condensed after the transition. Therefore, we expect the transition to be driven by the fluctuation of the order parameter $V_C$. The $B$ anyon, on the other hand, is not condensed. Therefore, we set $\phi_b =0$ and sum over the terms involving $\psi_c$. The resulting Lagrangian is
\begin{equation}
\begin{array}{lll}
\mathcal{L}_C & = &  K_{\tau} |\partial_{\tau}\psi_c|^2 + K_x |\partial_x \psi_c|^2 + \\
& & a |\psi_c|^2 + b (\psi_c^3 + \bar{\psi}_c^3) + c |\psi_c|^4 + ...
\end{array}
\end{equation}
as expected. In particular, the chiral term $\phi_b \left(\psi_c\partial_x\bar{\psi}_c-\bar{\psi}_c\partial_x\psi_c\right)$ is not allowed at this transition and the critical point is the normal 3-state Potts CFT.

The transition from the $S_3$ to $Z_2$ symmetry-breaking phase to the fully symmetry-breaking phase (bottom boundary $\mathcal{A}_3$ to $\mathcal{A}_1$) is more interesting. The $C$ anyon is condensed on both sides of the transition, hence $\psi_c$ is nonzero in the field theory. Before the transition in $\mathcal{A}_3$, $C$ is condensed in one of the reflection symmetric directions and we set $\psi_c$ to take the corresponding value in the Lagrangian. Suppose $C$ is condensed in the direction of $\ket{c}+\ket{\bar{c}}$, then $\psi_c = \bar{\psi}_c$ takes a nonzero real value. The $B$ anyon is condensed only after the transition, hence $\phi_b$ is the dynamical field driving the transition. Compared to the first case ($S_3$ symmetric to $S_3$ to $Z_3$ transition), more terms in section~\ref{sec:invertible_Op} are allowed, for example
\[
|\psi_c|^2, |\psi_c|^4, |\psi_c|^2\phi_b^2
\]
These terms are either constants ($|\psi_c|^2$, $|\psi_c|^4$) or reduce to existing terms of $\phi_b$ ($|\psi_c|^2\phi_b^2$). In particular, terms like $\phi_b \left(\psi_c\partial_x\bar{\psi}_c-\bar{\psi}_c\partial_x\psi_c\right)$ and $\phi_b\left(\psi_c^3-\bar{\psi}_c^3\right)$ are always zero because $\psi_c$ is real. There is no first-order term of $\phi_b$ that breaks the $Z_2$ symmetry across the transition. The Lagrangian describing the transition is still $\mathcal{L}_B$. After the transition, $\phi_b$ becomes nonzero. The chiral term changes the potential of $\psi_c$ and shifts the condensed direction of $C$. With a nonzero $\phi_b$, the potential of $C$ breaks the reflection symmetry and $C$ can now condense in an arbitrary direction in the two-dimensional internal space, hence the factor of $2$ in $\mathcal{A}_1$.

The transition from $S_3$ to $Z_3$ symmetry-breaking phase to the fully symmetry-breaking phase (bottom boundary $\mathcal{A}_2$ to $\mathcal{A}_1$) is an interesting one. The $B$ anyon is condensed on both sides, therefore $\phi_b$ can be set to be a real constant in the Lagrangian. The $C$ anyon becomes condensed after the transition, therefore $\psi_c$ is the field that drives the transition. All the terms listed in section~\ref{sec:invertible_Op} can be nonzero, but some of them are constants (like $\phi_b^2$). Combining all the dynamical terms (those involving $\psi_c$), we get the Lagrangian
\[
\mathcal{L}_{C,B} = \mathcal{L}_C + b_1 \phi_b \left(\psi_c\partial_x\bar{\psi}_c-\bar{\psi}_c\partial_x\psi_c \right) + ...
\]
In particular, the chiral term $\phi_b \left(\psi_c\partial_x\bar{\psi}_c-\bar{\psi}_c\partial_x\psi_c \right)$ is now activated and the field theory describes the chiral $Z_3$ transition\cite{Huse1982,Chepiga2019,Maceira2022}.

When the bottom boundary changes from $\mathcal{A}_3$ to $\mathcal{A}_2$, the system transitions from the $S_3$ to $Z_2$ symmetry-breaking phase to the $S_3$ to $Z_3$ symmetry-breaking phase. The symmetries on the two sides of the transition are not compatible and the transition is not a symmetry-breaking transition. 

When the bottom boundary changes from $\mathcal{A}_4$ to $\mathcal{A}_1$, the system transitions from the fully symmetric phase to the fully symmetry-breaking phase. Both $\phi_b$ and $\psi_c$ are dynamical fields across the transition and the Lagrangian contains all the possible terms listed in section~\ref{sec:invertible_Op}. It is not clear what kind of critical point is at the transition and we will not try to analyze the resulting Lagrangian.


\section{Noninvertible Ginzburg-Landau}
\label{sec:non-invertible}

With the preparation in the previous section, we are now ready to explain how to generalize the Ginzburg-Landau procedure to the symmetry-breaking transitions of non-invertible symmetries. 

\subsection{The Generalized Landau setup}
\label{sec:non-invertible_SymTFT}

First, we explain briefly why the symmetry-breaking transition of non-invertible symmetries is an important subject to study. This is because, barring some exceptions\cite{Pollmann2012a,Kobayashi2026,Warman2026,Gai2026}, almost all transitions between gapped phases of finite symmetries in $1+1$D can be mapped to symmetry-breaking transitions of non-invertible symmetries. 

\begin{figure}[th]
\begin{center}
\includegraphics[width= 1.0\textwidth]{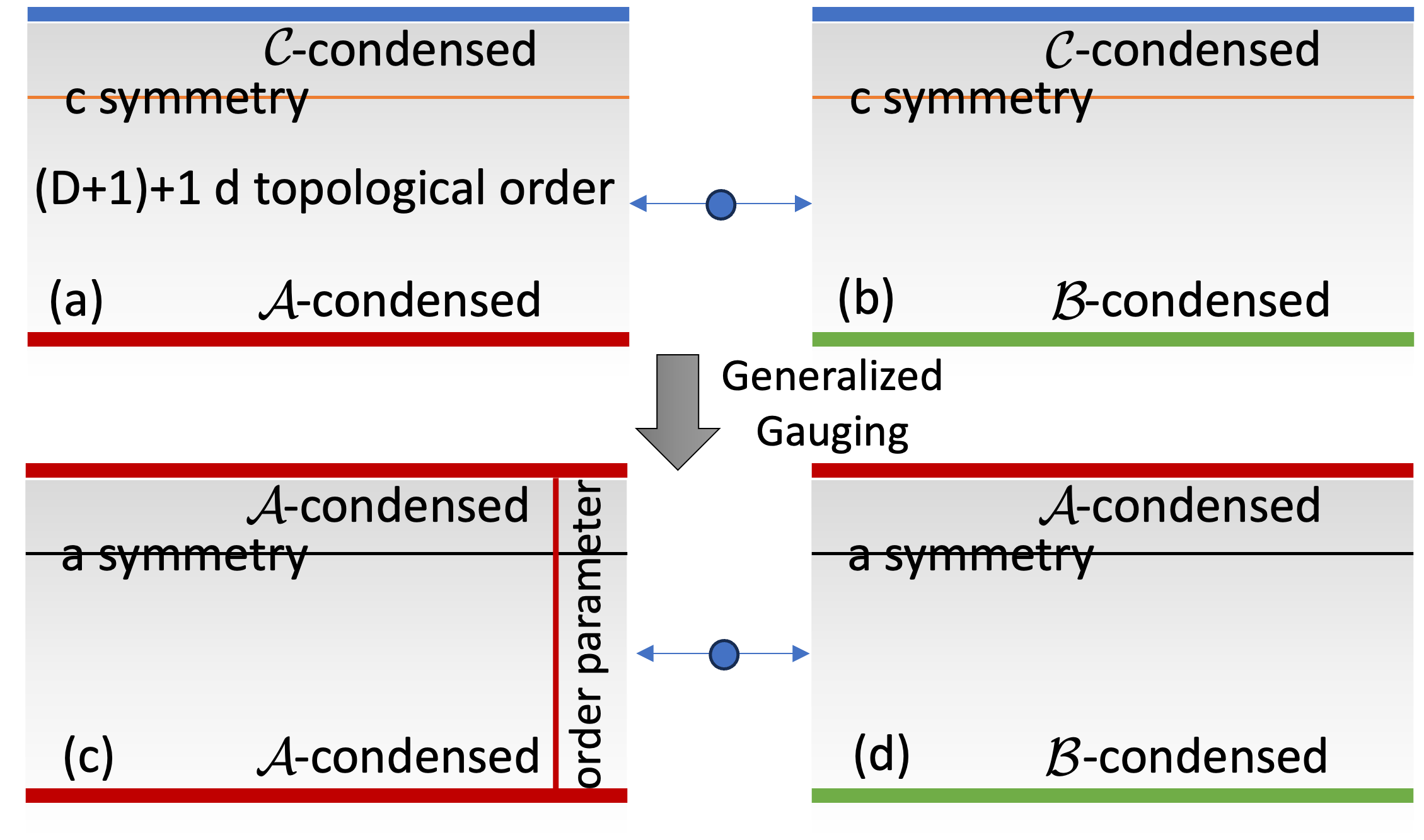}
\caption{Generalized gauging (changing the top boundary from $\mathcal{C}$ to $\mathcal{A}$) maps general gapped phases (a) to fully symmetry-breaking phases (c) and the phase transition between general gapped phases (a) and (b) to symmetry-breaking transitions between (c) and (d). Blue dots indicate critical transition points between gapped phases. The red line represents the tunneling of anyons between matching condensates on the two boundaries which becomes the order parameter in the spontaneous breaking of $a$-symmetry.} 
\label{fig:gengauge}
\end{center}
\end{figure}

Ref.~\cite{Chen2025} explained how such a mapping works through a generalized gauging procedure. We review this procedure here. 
Consider a $1+1$D gapped phase realized in a SymTFT sandwich structure with a $2+1$D topological order in the bulk, a gapped boundary at the top described by the Lagrangian algebra $\mathcal{C}$ and a gapped boundary at the bottom described by the Lagrangian algebra $\mathcal{A}$, as shown in the panel (a) of Fig.~\ref{fig:gengauge}. Such a phase can be mapped to a fully symmetry-breaking phase through a generalized gauging procedure where the top boundary of the sandwich is changed from $\mathcal{C}$ to $\mathcal{A}$, changing the symmetry of the sandwich from $c$-symmetry to $a$-symmetry. This is illustrated on the left hand side of Fig.~\ref{fig:gengauge} from panel (a) to panel (c). In general, the $a$-symmetry can be non-invertible if $\mathcal{A}$ does not correspond to the irreps of a group. After the mapping, all the anyons condensed in $\mathcal{A}$ can tunnel from the top to the bottom boundary through the red vertical operators in Fig.~\ref{fig:gengauge} without creating any excitations. These vertical operators are charged under the $a$-symmetry and become the order parameters of the $a$-symmetry. Therefore, the phase after the mapping is one where the $a$-symmetry is fully broken. Hence all gapped phases (that can fit into a sandwich) can be gauged (in a generalized sense) into fully symmetry-breaking phases, with potentially noninvertible symmetries. This was pointed out in Ref.~\cite{Lootens2024} using the tensor network formalism and was used to simplify numerical simulations of $1+1$D gapped models. 

Now, consider a second gapped phase realized in the same sandwich structure (the same bulk and the same top boundary labeled by $\mathcal{C}$) but with a $\mathcal{B}$-condensed bottom boundary (panel (b) of Fig.~\ref{fig:gengauge}). Applying the same generalized gauging procedure changes the top boundary to $\mathcal{A}$ (shown on the right hand side of Fig.~\ref{fig:gengauge} from panel (b) to panel (d)). The phase in panel (c) is fully symmetry-breaking, while the phase in panel (d) is not because fewer anyons can tunnel between the top and bottom boundaries (exceptions explained below). Therefore, the transition between the two original phases (panel (a) and (b)) is mapped to a symmetry-breaking transition between the phases in panel (c) and (d). This is illustrated in Fig.~\ref{fig:gengauge} from the top two panels to the bottom two panels.  The symmetry-breaking transition between (c) and (d) is induced by the fluctuation of the order parameter of the generalized $a$-symmetry, a phenomenon that would be described by a properly generalized Ginzburg-Landau theory.

For example, consider the sandwich structure with the bulk being the Quantum Double of $S_3$ and the top boundary being $\mathcal{C}=\mathcal{A}_1 = A+B+2C$. When the bottom boundary changes from $\mathcal{B} = \mathcal{A}_2 = A+B+2F$ to $\mathcal{A} = \mathcal{A}_3 = A+C+D$, the system transitions from the $S_3$ to $Z_3$ symmetry-breaking phase to the $S_3$ to $Z_2$ symmetry-breaking phase, which is not a symmetry-breaking transition. Changing the top boundary of the sandwich to $\mathcal{A} = \mathcal{A}_3$ corresponds to gauging the $Z_2$ subgroup of the $S_3$ symmetry. The symmetry of the system is now generated by $W_B$ and $W_F$, which satisfy 
\[
W_B^2 = W_A, W_F^2 = W_A+W_B+W_F, W_BW_F = W_F
\]
This is the non-invertible symmetry of $\mathrm{Rep}(S_3)$. Now when the bottom boundary changes from $\mathcal{B} = \mathcal{A}_2 = A+B+2F$ to $\mathcal{A} = \mathcal{A}_3 = A+C+D$, the system goes from the fully symmetric phase to the fully symmetry-breaking phase of $\mathrm{Rep}(S_3)$. The transition becomes a symmetry-breaking transition, of the non-invertible $\mathrm{Rep}(S_3)$ symmetry. More examples of generalized gauging mapping non-symmetry-breaking transitions to symmetry-breaking transitions include the Kennedy-Tasaki transformation\cite{Kennedy1992a,Kennedy1992b} mapping the transition between the $Z_2\times Z_2$ SPT phases to the $Z_2\times Z_2$ symmetry-breaking transition\cite{Huang2023topological}, and the transformation that maps the deconfined critical point between partial symmetry-breaking phases of the anomalous $Z_2\times Z_2$ symmetry to the symmetry-breaking transition of $Z_4$ symmetry\cite{Zhang2023}.

There is a class of examples for which this procedure is not effective, first discussed in Ref.~\cite{Pollmann2012a, Kobayashi2026} and dubbed twin phases in Refs.~\cite{Warman2026,Gai2026}. One example is associated with the group $G=Q_8\rtimes S_3$ of size 48 \cite{Warman2026,Gai2026}. In the SymTFT setup with the bulk topological order as the quantum double of $G$, the twin phases correspond to two distinct gapped bottom boundaries which condense the same set of anyons but differ in the associated multiplication map $M$. Under generalized gauging, the same set of anyons can tunnel into both boundaries and these two phases are both mapped to fully symmetry-breaking phases of the same non-invertible symmetry, yet they are distinct phases with different algebras of order parameters. If we were to directly apply the generalized Ginzburg-Landau analysis as prescribed above, there is no order parameter to assign dynamical fields to.         

\subsection{Order Parameter algebra}
\label{sec:non-invertible_Op}

Now with this setup, we will explain how to formulate the Ginzburg-Landau field theory for the symmetry-breaking transitions of non-invertible symmetries in $1+1$D. We will always consider situations where the symmetry can be realized in a SymTFT sandwich structure with a bulk $2+1$D topological order and a gapped top boundary $\mathcal{A}$, and the transition happens near the bottom boundary which changes from one gapped state $\mathcal{B}$ to the gapped state $\mathcal{A}$. The transition is hence always into a fully symmetry-breaking phase. The procedure works only in this setup. If neither phase is fully symmetry-breaking, the procedure can give the wrong result. We will illustrate each step of the procedure using the non-invertible symmetry $\mathrm{Rep}(S_3)$.  

In the general SymTFT setup with a bulk topological order and a gapped top boundary described by a Lagrangian algebra $\mathcal{A} = \sum_{\beta} n_{\beta} \beta$, the order parameters of the generalized symmetry are given by the vertical string operators $V_{\beta}$. We will consider only the situation where the $\beta$ anyons have integer quantum dimensions $d_{\beta}$ in the bulk. In the discussion section (section~\ref{sec:discussion}), we will comment on the possibility of generalizing to non-integer $d_{\beta}$. 

To realize the $\mathrm{Rep}(S_3)$ symmetry, we use the sandwich structure with the $S_3$ Quantum Double topological order in the bulk, but with a different boundary at the top. Among $\mathcal{A}_1$ through $\mathcal{A}_4$, $\mathcal{A}_1$ and $\mathcal{A}_2$ give rise to the $S_3$ symmetry while $\mathcal{A}_3$ and $\mathcal{A}_4$ give rise to the $\mathrm{Rep}(S_3)$ symmetry. For the following discussion, we will choose $\mathcal{A}_3 = A+C+D$ to be the top boundary. In Appendix~\ref{app:A4}, we discuss how the case of $\mathcal{A}_4 = A+D+F$ as the top boundary is similar and different from the case of $\mathcal{A}_3$. With $\mathcal{A}_3$ as the top boundary, the $\mathrm{Rep}(S_3)$ symmetry is generated by horizontal string operators $W_{B}$, $W_{F}$. The symmetry is non-invertible because $W_B$ and $W_F$ satisfy
\[
W_B^2 = W_A, W_F^2 = W_A+W_B+W_F, W_BW_F = W_F
\]
which is exactly the fusion rule of the irreps of $S_3$. The order parameters are given by vertical string operators $V_C$ and $V_D$. In a spin chain, the $\mathrm{Rep}(S_3)$ symmetry can be obtained by gauging a $Z_2$ subgroup of the $S_3$ symmetry. After gauging, the $B$ type order parameter of $S_3$ is replaced by a $D$ type order parameter, which represents the $Z_2$ flux sector of $S_3$. We will study the transitions where the bottom boundary changes from $\mathcal{A}_1$, $\mathcal{A}_2$, or $\mathcal{A}_4$ to $\mathcal{A}_3$. 

As explained in Section~\ref{sec:invertible_SymTFT}, the bottom index of $V_{\beta}$ labels the dynamical degrees of freedom that transform under the (generalized) symmetry. We need to assign dynamical fields to these dimensions. The rules for assigning dynamical fields to irreps, as explained in section~\ref{sec:invertible_review}, can be generalized to order parameters labeled by anyons $\beta$. 
\begin{enumerate}[label=\alph*)]
\item If $\beta$ is not its own anti-particle (two $\beta$'s cannot fuse into identity), we should assign complex fields to each of the internal dimensions of $\beta$ (labeled by the bottom index of $V_{\beta}$). 
\item If $\beta$ is its own anti-particle and two $\beta$'s fuse into identity in a way that is anti-symmetric under the exchange of the two $\beta$'s, we need to assign complex fields to the internal dimensions of $\beta$ but the fields are constrained.
\item If $\beta$ is its own anti-particle and two $\beta$'s fuse into identity in a way that is symmetric under the exchange of the two $\beta$'s, we can assign real fields to the internal dimensions of $\beta$. The basis of $\beta$ to which real fields are assigned can be determined from the requirement that independent Lagrangian terms are real. 
\end{enumerate}

Let's see how these rules apply to the order parameters $V_C$ and $V_D$ of $\mathrm{Rep}(S_3)$. Both $C$ and $D$ are their own anti-particle and both fuse into identity in a symmetric way. The identity fusion channel of two $C$'s corresponds to the singlet state
\[
\ket{c\bar{c}} + \ket{\bar{c}c}
\]
The identity fusion channel of two $D$'s corresponds to the singlet state
\[
\ket{d_1d_1}+\ket{d_2d_2}+\ket{d_3d_3}
\]
As discussed in section~\ref{sec:invertible_Op}, the identity fusion channels correspond to invariant terms in the Lagrangian. From the requirement that independent Lagrangian terms are real, we can determine how to assign fields to the components of $C$ and $D$. Suppose that $\psi_c$ and $\psi_{\bar{c}}$ are assigned to the dimensions $\ket{c}$ and $\ket{\bar{c}}$ of $C$ respectively. The two-body singlet state corresponds to the order two Lagrangian term $\psi_c\psi_{\bar{c}}$. For this term to be real, we arrive at the familiar requirement that $\psi_{\bar{c}} = \bar{\psi}_c$. For $D$, suppose that $\phi_{d_k}$'s are assigned to the dimensions $\ket{d_k}$. The two-body singlet term corresponds to the order two Lagrangian term $\sum_k \phi_{d_k}^2$. For this term to be real, all three of $\phi_{d_k}$'s are real. 

Note that the $\phi_{d_k}$'s are not local with respect to each other, because the dimensions $\ket{d_k}$ braid nontrivially with each other. We should add topological terms to the Lagrangian to account for such non-locality when $\phi_{d_k}$'s appear together. As we will see, at a $\mathrm{Rep}(S_3)$ symmetry breaking transition, only one of the $\phi_{d_k}$'s drives the transition and therefore we don't need to worry about their mutual non-locality if we focus on the fluctuation of only one of them.

Having assigned field variables to the components of the order parameters and discussed how they form order two invariant Lagrangian terms, we can start to form invariant terms of higher order in the Lagrangian. Such terms correspond to the identity fusion trees (illustrated in Fig.~\ref{fig:FusionTree}) with more branches. Before discussing the third and higher order terms, we need to spend some time to look at how the top boundary constrains the possibilities of allowed fusion trees when the symmetry of the sandwich structure is non-invertible. 

In section~\ref{sec:invertible_Op}, we argued that only the bottom part of the fusion tree is important while the top part only adds an unimportant prefactor to each term. This conclusion holds for invertible symmetries. For non-invertible symmetries, the top part of the fusion tree is also important because the prefactor it adds can be zero, effectively constraining the possible terms. 

To see where the constraints come from, let's look more closely at the Lagrangian algebra $\mathcal{A}_3 = A+C+D$. In $\mathcal{A}_3 = A+C+D$, both $C$ and $D$ condense with only one internal dimension. In section~\ref{sec:invertible_transition}, we explained that $C$ condenses in one of three directions $\ket{c}+\ket{\bar{c}}$, $\ket{c}+\omega\ket{\bar{c}}$, or $\ket{c}+\bar{\omega}\ket{\bar{c}}$. The condensation of $D$ needs to be compatible with the condensation of $C$. That is, the condensable dimension of $D$ needs to commute with the condensable dimension of $C$. The commutation between an internal dimension of a pure gauge charge ($B$ or $C$) and that of a pure flux ($D$ or $F$) is given by\cite{Brennen2009}
\begin{equation}
\mathcal{R}^2 \ket{\rho_i}\ket{g} = \sum_j T^{(\rho)}_{ij}(g)\ket{\rho_j}\ket{g} 
\label{eq:Rcf}
\end{equation}
where $\mathcal{R}$ denotes the exchange of the two states and $\mathcal{R}^2$ the full braid, $\ket{\rho_i}$ is the $i$th dimension of the gauge charge labeled by irrep $\rho$, $\ket{g}$ is the flux dimension labeled by group element $g$, and $T^{(\rho)}$ are the representing matrices of irrep $\rho$. Using this equation, we see that $D$ needs to condense in either $\ket{d_1}$, $\ket{d_2}$ or $\ket{d_3}$ to be compatible with the condensation of $C$. On the other hand, the commutation between the internal dimensions of two pure fluxes is
\begin{equation}
\mathcal{R} \ket{g}\ket{h} = \ket{h}\ket{h^{-1}gh}
\label{eq:Rff}
\end{equation}
Hence the condensation of individual $\ket{d_k}$'s is compatible by itself. Therefore, for $\mathcal{A}_3$, we find three sets of condensable dimensions
\[
\begin{array}{ll}
C: \frac{1}{\sqrt{2}}\left(\ket{c}+\ket{\bar{c}}\right) & D: \ket{d_1} \\
C: \frac{1}{\sqrt{2}}\left(\ket{c}+ \bar{\omega}\ket{\bar{c}}\right) & D: \ket{d_2} \\
C: \frac{1}{\sqrt{2}}\left(\ket{c}+\omega\ket{\bar{c}}\right) & D: \ket{d_3} \\
\end{array}
\]
Each set of condensable dimensions forms an `algebra' meaning that their multiplication is closed. The multiplication coefficient $M$ (see Fig.~\ref{fig:OpAl}) can be found from the following calculation (done for the first set of condensable dimensions).
\[
\begin{array}{ll}
& C \times_{\mathcal{A}_3} C \\
= & \frac{1}{\sqrt{2}} \left(\ket{c}+\ket{\bar{c}}\right)\frac{1}{\sqrt{2}} \left(\ket{c}+\ket{\bar{c}}\right) \\
= & \frac{1}{2}\left(\ket{c\bar{c}}+\ket{\bar{c}c}+\ket{cc}+\ket{\bar{c}\bar{c}} \right) \\ 
\to & \frac{1}{\sqrt{2}} A + \frac{1}{\sqrt{2}} C
\end{array}
\]
where $\times_{\mathcal{A}_3}$ denotes the multiplication of condensable dimensions on $\mathcal{A}_3$. This is very different from the fusion of two $C$'s in the bulk. In the bulk, $C\times C = A+B+C$. But on $\mathcal{A}_3$, two $C$'s do not multiply into $B$. Therefore, this fusion channel is not allowed in the fusion tree of Fig.~\ref{fig:FusionTree} and the corresponding term is not allowed in the Lagrangian. 

Similarly, the fusion of two $D$'s on $\mathcal{A}_3$ is very different from that in the bulk.
\[
\begin{array}{ll}
& D \times_{\mathcal{A}_3} D \\
= & \ket{d_1}\ket{d_1} \\
= & \frac{1}{3} \left(\ket{d_1d_1}+\ket{d_2d_2}+\ket{d_3d_3}\right) + \\
& \frac{1}{3} \left(\ket{d_1d_1}+\omega\ket{d_2d_2}+\bar{\omega}\ket{d_3d_3}\right) + \\
& \frac{1}{3} \left(\ket{d_1d_1}+\bar{\omega}\ket{d_2d_2}+\omega\ket{d_3d_3}\right) \\
\to & \frac{1}{\sqrt{3}}A + \frac{\sqrt{2}}{\sqrt{3}} C
\end{array}
\]
Therefore, even though in the bulk $D\times D = A+C+F+G+H$, the $F$, $G$, $H$ fusion channels are not allowed in the fusion tree of Fig.~\ref{fig:FusionTree} and the corresponding terms are not allowed in the Lagrangian. 

Finally, the fusion of $C$ and $D$ on $\mathcal{A}_3$ goes as
\[
\begin{array}{ll}
& C \times_{\mathcal{A}_3} D \\
= & \frac{1}{\sqrt{2}} \left(\ket{c}+\ket{\bar{c}}\right)\ket{d_1} \\
\to & D
\end{array}
\]
Therefore, even though in the bulk $C\times D = D+E$, the $E$ fusion channel is not allowed in the fusion tree of Fig.~\ref{fig:FusionTree}.

We see that the effect of the top boundary $\mathcal{A}_3$ is to constrain the branches of the fusion tree to be within the set of anyons that label order parameters. This conclusion holds for all SymTFT sandwich structures. Fusion trees with branches outside of the set must be zero. For the allowed fusion trees where all branches are within the set, the top boundary simply adds a prefactor and does not affect the resulting Lagrangian term.

Taking this constraint into account, we can now list all possible Lagrangian terms. We will do this up to the fourth order. Note that while the top indices of the order parameters ($\mu$, $\nu$ in Fig.~\ref{fig:OpAl}) are fixed due to the condensation in $\mathcal{A}_3$, the bottom indices are open to fluctuation.

At second order, there are two possible fusion trees,

\begin{figure}[h]
\begin{center}
\includegraphics[width= 0.4\textwidth]{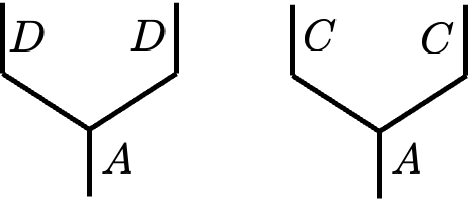}
\end{center}
\end{figure}

which correspond to quantum states
\[
\ket{c\bar{c}}+\ket{\bar{c}c}, \ \ket{d_1d_1}+\ket{d_2d_2}+\ket{d_3d_3}
\]
Replacing the operators with their assigned fields, we obtain the quadratic Lagrangian terms
\[
|\psi_c|^2, \phi_{d_1}^2+\phi_{d_2}^2+\phi_{d_3}^2
\]
If we allow for spatial and time derivatives (assuming that the $\mathrm{Rep}(S_3)$ symmetry is an internal, not a spatial, symmetry), we obtain terms like
\[
 |\partial_{\tau} \psi_c|^2, |\partial_x \psi_c|^2, \sum_{k}\left(\partial_{\tau} \phi_{d_k}\right)^2, \sum_{k}\left(\partial_x \phi_{d_k}\right)^2,
\]

At third order, there are two possible fusion trees,

\begin{figure}[h]
\begin{center}
\includegraphics[width= 0.6\textwidth]{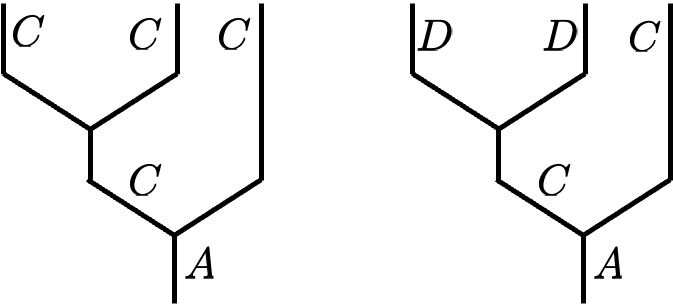}
\end{center}
\end{figure}

which correspond to quantum states
\[
\begin{array}{l}
\ket{ccc}+\ket{\bar{c}\bar{c}\bar{c}}, \\ \left(\ket{d_1d_1}+\omega\ket{d_2d_2}+\bar{\omega}\ket{d_3d_3}\right)\ket{c}+ \\
\left(\ket{d_1d_1}+\bar{\omega}\ket{d_2d_2}+\omega\ket{d_3d_3}\right)\ket{\bar{c}}
\end{array}
\]
The corresponding Lagrangian terms are
\begin{equation}
\begin{array}{l}
\psi_c^3+\bar{\psi}_c^3, \\
\left(\phi_{d_1}^2+\omega\phi_{d_2}^2+\bar{\omega}\phi_{d_3}^2\right)\psi_c+
\left(\phi_{d_1}^2+\bar{\omega}\phi_{d_2}^2+\omega\phi_{d_3}^2\right)\bar{\psi}_c
\end{array}
\label{eq:rS3_3}
\end{equation}

At fourth order, we need to take into account the constraints coming from the top boundary. That is, only fusion trees with branches in the set $A$, $C$, $D$ are allowed. We can draw six such fusion trees, two with four $D$'s, two with four $C$'s and two with two $C$'s and two $D$'s.

\begin{figure}[h]
\begin{center}
\includegraphics[width= 0.5\textwidth]{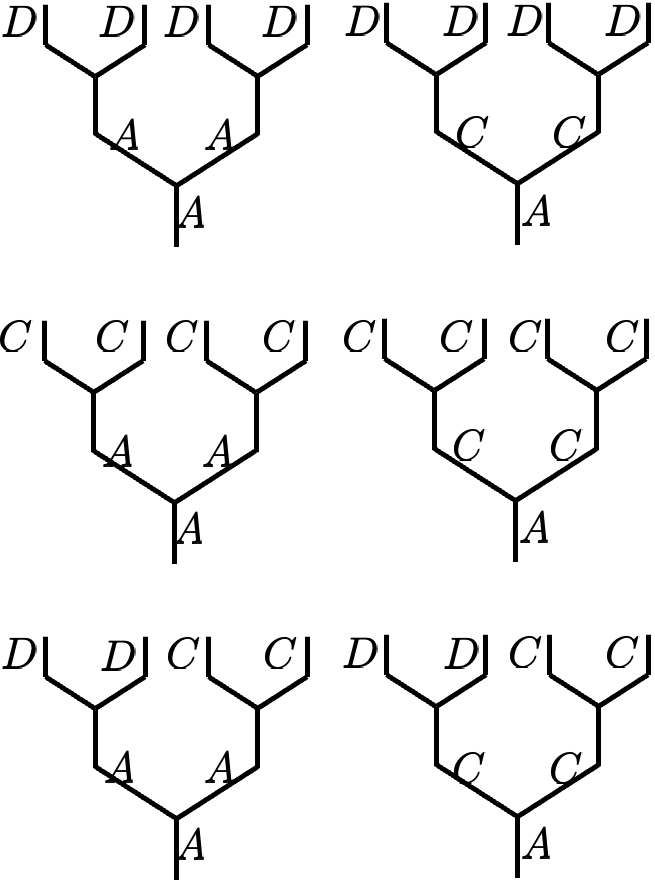}
\end{center}
\end{figure}

But there are actually more constraints. The fusion trees shown above are drawn according to a particular fusion order (1 fuses with 2, 3 fuses with 4, then their results fuse together). Changing the fusion order (for example to 1 fuses with 2, then with 3, then with 4) corresponds to a linear transformation among the fusion trees and might result in trees not allowed by the top boundary. To ensure that the fusion tree is always within the allowed space, we make linear superpositions as shown in the following figure, such that the trees do not contain forbidden branches no matter what the fusion order is.

\begin{figure}[h]
\begin{center}
\includegraphics[width= 0.5\textwidth]{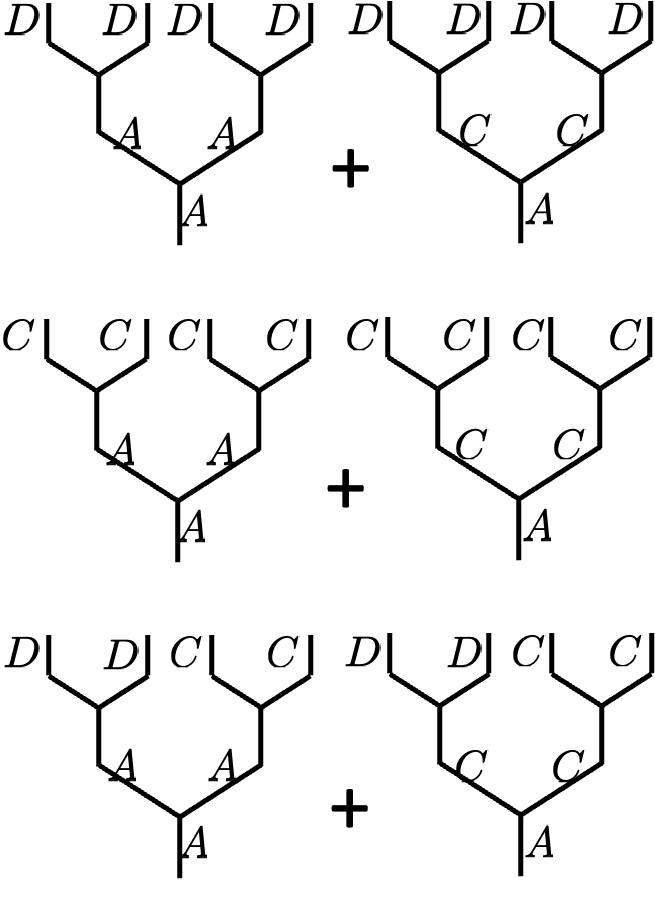}
\end{center}
\end{figure}

To see this, we write down the quantum states corresponding to the superposed trees
\[
\begin{array}{l}
\ket{d_1d_1d_1d_1}+\ket{d_2d_2d_2d_2}+\ket{d_3d_3d_3d_3}, \\
\ket{cc\bar{c}\bar{c}} + \text{permutation}, \\
\left(\ket{d_1d_1}+\ket{d_2d_2}+\ket{d_3d_3}\right)\left(\ket{c\bar{c}}+\ket{\bar{c}c}\right) \\
+ \left(\ket{d_1d_1}+\omega\ket{d_2d_2}+\bar{\omega}\ket{d_3d_3}\right)\ket{\bar{c}\bar{c}} \\
+ \left(\ket{d_1d_1}+\bar{\omega}\ket{d_2d_2}+\omega\ket{d_3d_3}\right)\ket{cc}
\end{array}
\]
The first two states (corresponding to fusion trees with four $C$'s and four $D$'s) are obviously permutation invariant.  For the last state, it can also be checked that the fusion branches are always allowed for any fusion order. 

As a result, the allowed Lagrangian terms are
\[
\begin{array}{l}
\phi_{d_1}^4 + \phi_{d_2}^4 + \phi_{d_3}^4, \\
|\psi_c|^4, \\
\left(\phi_{d_1}^2 + \phi_{d_2}^2 + \phi_{d_3}^2\right)|\psi_c|^2 \\
+ \left(\phi_{d_1}^2 + \omega\phi_{d_2}^2 + \bar{\omega}\phi_{d_3}^2\right)\bar{\psi_c}^2 \\
+ \left(\phi_{d_1}^2 + \bar{\omega}\phi_{d_2}^2 + \omega\phi_{d_3}^2\right)\psi_c^2
\end{array}
\label{eq:rS3_4}
\]

\subsection{Field Theory Description of Transitions}
\label{sec:non-invertible_transition}

Using the Lagrangian terms found in the last section, we can construct the Ginzburg-Landau field theory describing the symmetry-breaking transition of $\mathrm{Rep}(S_3)$ symmetry. As explained in section~\ref{sec:non-invertible_SymTFT}, we will only consider transitions into the fully symmetry-breaking phase. In appendix~\ref{app:nonfull}, we discuss a transition of $\mathrm{Rep}(S_3)$ where this condition is not satisfied and show that the procedure discussed here does not recover the expected critical point. 
Lattice realizations of $\mathrm{Rep}(S_3)$-symmetric gapped phases and transitions between them have been studied in Ref.~\cite{Chatterjee2024,Bhardwaj2025,Bhardwaj2024,Bhardwaj2026}.

With the top boundary set at $\mathcal{A}_3 = A+C+D$, let's consider first the case where the bottom boundary changes from $\mathcal{A}_1=A+B+2C$ to $\mathcal{A}_3=A+C+D$. Before the transition, the $C$ order parameter is condensed, but $D$ is not. Since $V_C$ transforms under $W_F$ but not $W_B$, the $W_F$ symmetry is broken but $W_B$ is not. After the transition, the $\mathrm{Rep}(S_3)$ symmetry is fully broken. Therefore, the transition is a $Z_2$ symmetry-breaking transition. This can also be seen by changing the top boundary to $\mathcal{A}_1$, i.e. un-gauge the system back into one with $S_3$ symmetry. With $S_3$ symmetry, the transition is the Ising $Z_2$ transition driven by the $B$ type order parameter. Changing the top boundary back to $\mathcal{A}_3$ corresponds to gauging the $Z_2$ symmetry and the Ising critical point remains the same critical theory. Let's see if we recover this result from the constructed field theory.

To write down the field theory, we identify the different roles played by the $C$ and $D$ type order parameters. $C$ is condensed on both sides of the transition. Before the transition, it is condensed in a generic direction in the order parameter space (due to the coefficient $2$ in $\mathcal{A}_1$). As a result, we set $\psi_c$ to a generic complex value in the Lagrangian. It is a non-dynamical degree of freedom. $D$, on the other hand, becomes condensed after the transition. Therefore, it represents the dynamical degree of freedom driving the transition. 

At second order, the Lagrangian terms relevant for the transition are
\[
\sum_k\left(\partial_{\tau} \phi_{d_k}\right)^2, \sum_k\left(\partial_x \phi_{d_k}\right)^2, \sum_k \phi_{d_k}^2
\]
Having all three $\phi_{d_k}$ fields in the Lagrangian looks problematic because they are not local fields with respect to each other. In principle, we should include the proper topological term to account for their mutual non-locality. In fact, we can avoid this problem because a third order term breaks the degeneracy of the three fields so that their quadratic terms can have different coefficients. Consider the second third order term in Eq.~\ref{eq:rS3_3}. When $\psi_c$ takes on a generic complex value, this reduces effectively to a quadratic term for the $\phi_{d_k}$'s with different (real) coefficients. Therefore, when the coefficients in front of the quadratic terms vary and drive the transition, one of the $\phi_{d_k}$'s makes the transition while the other two remain gapped. Therefore, in the Lagrangian, we only need to include the dynamical terms of one of the $\phi_{d_k}$'s. Which one is relevant for the transition depends on the choice of $\psi_c$. If we assume WLOG that $\phi_{d_1}$ is the one that drives the transition, the Lagrangian contains terms like
\[
\begin{array}{lll}
\mathcal{L} & = & K_{\tau} \left(\partial_{\tau}\phi_{d_1}\right)^2 + K_x \left(\partial_x \phi_{d_1}\right)^2 + \\
& & a \phi_{d_1}^2 + b \phi_{d_1}^4 + ...
\end{array}
\]
Obviously this is the same Lagrangian as $\mathcal{L}_B$, which describes the Ising $Z_2$ transition, as expected. 

Another interesting feature of this transition is that the condensable dimension of $C$ reduces from 2 to 1 after the transition. That is, before the transition, the order parameter $C$ can point in a generic direction in the two-dimensional internal space, but after the transition it is restricted to point to isolated directions in the space. The restriction comes from the condensation of $D$. Suppose that after the transition, $D$ is condensed in the $\ket{d_1}$ state, i.e. the coefficient in front of $\phi_{d_1}^2$ first becomes negative. When this happens, $C$ cannot point in any direction any more. To be consistent with the condensation of $D$, $C$ needs to adapt and move to the direction of $\ket{c}+\ket{\bar{c}}$. The mixed third order term becomes a linear potential term for $C$ and pins it in the $\psi_c = \bar{\psi}_c$ direction. Note that this change is the result of the condensation of $D$. $C$ is not the dynamical field that drives the transition. This is a phenomenon that happens only for non-invertible symmetries.  

Next, let's consider the case where the bottom boundary changes from $\mathcal{A}_4=A+D+F$ to $\mathcal{A}_3=A+C+D$. Before the transition, the $D$ order parameter is condensed but $C$ is not. As a result, the $W_B$ symmetry is broken. The $W_F$ symmetry splits into two parts and generates the remaining $Z_3$ symmetry. After the transition, the $\mathrm{Rep}(S_3)$ symmetry is fully broken. The transition is some kind of $Z_3$ symmetry-breaking transition. 

By un-gauging the $Z_2$ symmetry of $\mathrm{Rep}(S_3)$ (mapping the top boundary to $\mathcal{A}_1$), we see that the transition is from the $S_3$ symmetric phase to the $S_3$ to $Z_2$ symmetry-breaking phase, with the critical point being the 3-state Potts CFT. Therefore, when the $Z_2$ symmetry is gauged, we should get the tetra-critical Ising CFT with $\mathrm{Rep}(S_3)$ symmetry. We hence expect to get the same field theory as $\mathcal{L}_C$, which describes the 3-state Potts CFT. The distinction between tetra-critical Ising and 3-state Potts can be identified by carefully analyzing the symmetry sectors of the CFT.

The $C$ order parameter is the dynamical field driving the transition. The $D$ order parameter, on the other hand, is condensed on both sides and non-dynamical. The $\phi_{d_k}$ fields should be set to constant values in the field theory. To find out which constant values to assign to the $\phi_{d_k}$ fields, we need to understand how $D$ condenses before the transition in $\mathcal{A}_4$. Using the commutation relation between symmetry fluxes given in Eq.~\ref{eq:Rff}, we find the commuting, and hence condensed, dimensions of $D$ and $F$ to be
\begin{equation}
D: \ket{d_1}+\ket{d_2}+\ket{d_3}, \ F: \ket{f}+\ket{f^2}
\end{equation}
(There are actually three sets of commuting dimensions of $D$ and $F$. We explain the details in Appendix~\ref{app:A4}.) As a result, the $\phi_{d_k}$ fields take on equal real expectation values
\[
\phi_{d_1} = \phi_{d_2} = \phi_{d_3} = \phi_d
\]
Because of this, the mixed third order term (second term in Eq.~\ref{eq:rS3_3}) becomes zero. The mixed fourth order term (the third term in Eq.~\ref{eq:rS3_4}) reduces to the quadratic term of $\psi_c$, $|\psi_c|^2$.

Therefore, combining all the dynamical terms, we arrive at the Lagrangian
\begin{equation}
\begin{array}{lll}
\mathcal{L} & = &  K_{\tau} |\partial_{\tau}\psi_c|^2 + K_x |\partial_x \psi_c|^2 + \\
& & a |\psi_c|^2 + b (\psi_c^3 + \bar{\psi}_c^3) + c |\psi_c|^4 + ...
\end{array}
\end{equation}
This is the same Lagrangian as $\mathcal{L}_C$, which describes the non-chiral $Z_3$ symmetry-breaking transition with the 3-state Potts CFT at the critical point. The actual critical theory is the tetra-critical Ising CFT, which differs from the 3-state Potts only on kinematic aspects. The 3-state Potts CFT has $S_3$ symmetry while the tetra-critical Ising CFT has $\mathrm{Rep}(S_3)$ symmetry. Therefore, the field theory above actually has the tetra-critical Ising CFT at the critical point.

Finally, the transition from $\mathcal{A}_2$ to $\mathcal{A}_3$ involves the condensation of both $C$ and $D$. Both the $\psi_c$ and the $\phi_{d_k}$ fields are dynamical across the transition and the field theory can contain all the terms listed in section~\ref{sec:non-invertible_Op}. It is not clear what happens at the critical point, and we will not try to analyze the obtained field theory.


\section{Discussion}
\label{sec:discussion}

Using the example of $S_3$ and $\mathrm{Rep}(S_3)$ symmetries, we explained how the Ginzburg-Landau procedure for invertible symmetry-breaking transitions can be re-interpreted in the SymTFT formalism and subsequently generalized to be applicable to non-invertible symmetry-breaking transitions. Some key conditions and steps in the generalized procedure are:
\begin{itemize}
\item In the SymTFT formalism, the bulk topological order and the top boundary determine the symmetry.
\item The order parameters of the symmetry are labeled by the anyons condensed on the top boundary, with operator algebra given in Fig.~\ref{fig:OpAl}. 
\item Real / complex fields are assigned to the internal dimensions of the order parameters depending on the real / complex nature of the anyon.
\item Invariant Lagrangian terms of the order parameters correspond to fusion trees that fuse into the trivial anyon (Fig.~\ref{fig:FusionTree}).
\item Due to constraints from the top boundary, only fusion trees with all the branches in the order parameter anyon set are allowed.
\item A field theory can be formulated using the invariant terms to describe a transition when the bottom boundary changes from one gapped state to another.
\item The procedure works when the top boundary of the SymTFT matches the bottom boundary after the transitions. That is, the transition is always into the fully symmetry-breaking phase, which can be ensured through proper generalized gauging. 
\item An order parameter is set to a constant value if the labeling anyon is condensed in the bottom boundary on both sides of the transition.
\item An order parameter is the dynamical field driving the transition if the labeling anyon becomes condensed after the transition.
\item Different components of an order parameter may not be local with respect to each other, but only a mutually-local subset drives a transition.
\end{itemize}

This procedure can be applied to all kinds of transitions in the $1+1$D SymTFT setup. The transitions discussed in this paper are either conventional Ginzburg-Landau (symmetry-breaking of $S_3$) or can be obtained by gauging a discrete subgroup of such a conventional theory (symmetry-breaking of $\mathrm{Rep}(S_3)$). In an upcoming work, we will study symmetry-breaking transitions of the non-invertible symmetry Rep($H_8$), which cannot be mapped to conventional symmetry-breaking transitions through generalized gauging and are hence truly beyond the original Ginzburg-Landau framework. 

This procedure does not work when the mapping (generalized gauging) used in the generalized Landau setup (section~\ref{sec:non-invertible_SymTFT}) does not produce symmetry breaking transitions. That happens when the two bottom boundaries condense the same set of anyons. Such cases are found in Ref.~\cite{Pollmann2012a,Kobayashi2026} and are systematically classified in Ref.~\cite{Warman2026,Gai2026}. However the generalized Ginzburg-Landau theory is sensitive to the condensed dimensions of the anyons, hence could provide different theories for transitions out of the twin phases into the same phase.

It is well known that the Ginzburg-Landau theory has limited utility in $1+1$D. In higher dimensions (e.g. $> 3+1$D), the mean-field approximation yields quantitatively correct results for the scaling behavior at the critical point. But in lower dimensions, the coupling terms become relevant and solving the field theory becomes nontrivial. In this work, we are not attempting to provide new methods for analyzing strongly coupled field theories in $1+1$D. Rather, we identify the field theory that describes the critical point of interest, whose analysis will need to rely on other methods (for example numerical simulation or mapping to free fermion theories). 

What we manage to achieve is to show that, despite the exotic properties of non-invertible symmetries, Landau's philosophy that symmetry-breaking transitions are driven by the fluctuation of order parameters still applies and that the algebra of the order parameters determines the universality class of the critical point. More specifically, the field theory obtained using the generalized Ginzburg Landau procedure is expected to correctly describe the universal dynamical aspects of the transition, like the dynamical exponent, central charge of a CFT, correlation functions, etc. It captures what happens near the bottom boundary of the SymTFT sandwich. The kinetic aspects of the theory, on the other hand,  depend on the top boundary of the SymTFT sandwich. To recover the correct symmetry or symmetry sectors in the Hilbert space, a generalized gauging step might be needed (for example to recover the tetracritical Ising CFT from the 3-state Potts CFT). 

In this paper, we only consider cases where the order parameters are labeled by anyons with integer quantum dimensions in the bulk. We suspect that a modified procedure can be applied even when the anyons have non-integer quantum dimensions. The difficulty lies in the step of assigning quantum fields to the order parameters. If the anyon labeling an order parameter has a non-integer quantum dimension, it is unclear how to assign fields to different components of the order parameter. But as we can see from the case of the $\mathrm{Rep}(S_3)$ transition driven by the condensation of $D$ (top boundary $\mathcal{A}_3$, bottom boundary $\mathcal{A}_1$ to $\mathcal{A}_3$), not all components of the order parameters are important. Even though the $D$ anyon has dimension 3, only one of the components condenses across the transition while the other two remain gapped. In general cases, even though the condensing anyons may have non-integer dimensions in the bulk, only an integer-dimensional subspace can condense. This is determined by the structure of the Lagrangian Algebra describing the bottom boundary where the coefficient in front of each anyon in the algebra is always integer. Therefore, to properly capture a transition driven by the condensation of an anyon, we may only need to take into account an integer dimensional subspace even though the anyon has other dimensions. Another interesting case to compare with the condensation of $D$ is the condensation of $C$ (top boundary $\mathcal{A}_3$, bottom boundary $\mathcal{A}_4$ to
$\mathcal{A}_3$). $C$ has two internal dimensions and it condenses with only coefficient 1 in $\mathcal{A}_3$. But when formulating the field theory, it is important that both dimensions are taken into account (i.e. $\psi_c$ is complex) because both become critical fields at the transition. This is possible because the two dimensions of $C$ commute with each other and can fluctuate consistently while the same is not true among the dimensions of $D$. 

At this point, we do not have a complete formulation of the generalized Ginzburg-Landau procedure when the condensing anyons have non-integer quantum dimensions, but maybe by properly identifying the subspace of internal dimensions of the anyons responsible for driving a condensation transition, it would be possible to assign fields and form Lagrangian terms only using this subspace even though the anyon has other dimensions. Such an understanding can be useful in coming up with a field theory description of anyon condensation transitions between $2+1$D topological phases as well, where the key data describing the transition is again the condensable algebra of condensable dimensions of a subset of anyons. 



\begin{acknowledgments}
We are grateful for inspiring discussions with Lukasz Fidkowski, John McGreevy, Shinsei Ryu, Sergej Moroz, Arkya Chatterjee, Bowen Yang, Zhenghan Wang. V.R. is supported by the National Science Foundation Graduate Research Fellowship under Grant No. 2139433. 
X.C. is supported by the Simons collaboration on `Ultra-Quantum Matter'' (grant number 651438), the Simons Investigator Award (award ID 828078), the Institute for Quantum Information and Matter at Caltech (grant number PHY-2317110), the Walter Burke Institute for Theoretical Physics at Caltech and the Leinweber Forum for Theoretical Physics at Caltech. L.E. is supported by the Walter Burke Institute for Theoretical Physics at Caltech.
\end{acknowledgments}

\appendix

\section{Basic data of $S_3$ Quantum Double}
\label{app:S3}

For convenience, we review the basic data of the $S_3$ Quantum Double topological order in this section. As with all group Quantum Doubles, the anyons are divided into pure charge, pure flux and dyons. The pure charges $A,B$ and $C$ are the trivial, sign and $2d$ irreducible representations (irreps) of $S_3$ respectively. These satisfy the usual fusion rules for irreps:
\begin{equation}
\begin{split}
     B \times B &= A\\
     C \times C &= A + B + C \\
     B \times C &= C
\end{split}
\end{equation}
By choosing bases for these irreps we can identify explicitly each fusion result in the tensor product of the two irreps. 

The pure fluxes correspond to the conjugacy classes in $S_3$. Calling the reflection and rotation group elements $d,f$ respectively, there are two such classes: $\{d_1,d_2,d_3\}$ (labeled $D$) and $\{f,\bar{f}\}$ (labeled $F$). These anyons satisfy the fusion rules:
\begin{equation}
\begin{split}
     F \times F &= A + B +F\\
     D \times D &= A + C + F+G+H \\
     D \times F &= D+E
\end{split}
\end{equation}
Here $E,G,H$ are dyons, with $E$ carrying the same flux as $D$ and $G,H$ carrying the same flux as $F$. Similar to irreps, we can decompose tensor products of the $D$ and $F$ conjugacy classes into these fusion outcomes. For example in the first fusion rule listed above, the $A$ outcome is spanned by $\ket{f\bar{f}}+\ket{\bar{f}f}$ and the 2 dimensional $F$ outcome by $\ket{ff},\ket{\bar{f}\bar{f}}$.

There is a $Z_2$ automorphism of the Quantum Double which acts by swapping the 2 dimensional $C$ and $F$ anyons. This swap preserves fusion rules and braiding. For example, we can take it to map the internal states as: $\ket{c},\ket{\bar{c}}\to \ket{f},\ket{f^2}$. This automorphism also swaps the two Lagrangian algebras $A+D+F$ and $A+C+D$. Since the condensed dimension of $D$ is different in these two algebras (App.~\ref{app:A4}), this automorphism must act non-trivially on the internal states of the $D$ anyon as well. 

\section{$\mathrm{Rep}(S_3)$ symmetry with $\mathcal{A}_4$ as the top boundary}
\label{app:A4}

In Section~\ref{sec:non-invertible}, we consider the SymTFT sandwich with the $\mathcal{A}_3 = A+C+D$ top boundary and the resulting phase transitions with $\mathrm{Rep}(S_3)$ symmetry. The $\mathcal{A}_4 = A+D+F$ top boundary also gives rise to the $\mathrm{Rep}(S_3)$ symmetry. This is easy to see given the symmetry between the $C$ and $F$ anyon in the bulk\cite{}. Therefore, all the results derived with the $\mathcal{A}_3$ top boundary should directly apply with the $\mathcal{A}_4$ top boundary. Nonetheless, there are interesting features in the $\mathcal{A}_4$ top boundary and the associated Ginzburg-Landau procedure, which we explain in this section.

In $\mathcal{A}_4$, both $D$ and $F$ condense with one of their internal dimensions. Using the commutation relation between the internal dimensions of pure fluxes (Eq.~\ref{eq:Rff})
\begin{equation}
\mathcal{R} \ket{g}\ket{h} = \ket{h}\ket{h^{-1}gh}
\end{equation}
we see that one possible set of condensable dimensions for $D$ and $F$ are
\begin{equation}
D: \ket{d_1}+\ket{d_2}+\ket{d_3}, F: \ket{f}+\ket{\bar{f}} 
\end{equation}

These two dimensions, together with the dimension of the trivial anyon $A$, form the closed Lagrangian Algebra under multiplication. In particular, the product of two $F$'s goes like
\[
\begin{array}{ll}
& F \times_{\mathcal{A}_4} F \\
= & \frac{1}{\sqrt{2}} \left(\ket{f}+\ket{\bar{f}}\right)\frac{1}{\sqrt{2}} \left(\ket{f}+\ket{\bar{f}}\right) \\
= & \frac{1}{2}\left(\ket{f\bar{f}}+\ket{\bar{f}f}+\ket{ff}+\ket{\bar{f}\bar{f}} \right) \\ 
\to & \frac{1}{\sqrt{2}} A + \frac{1}{\sqrt{2}} F
\end{array}
\]
The product of two $D$'s goes like
\[
\begin{array}{ll}
& D \times_{\mathcal{A}_4} D \\
= & \frac{1}{\sqrt{3}}\left(\ket{d_1}+\ket{d_2}+\ket{d_3}\right)^{\otimes 2} \\
= & \frac{1}{3} \left(\ket{d_1d_1}+\ket{d_2d_2}+\ket{d_3d_3}\right) + \\
& \frac{1}{3} \left(\ket{d_1d_2}+\ket{d_2d_3}+\ket{d_3d_1}\right) + \\
& \frac{1}{3} \left(\ket{d_2d_1}+\bar{\omega}\ket{d_3d_2}+\omega\ket{d_1d_3}\right) \\
\to & \frac{1}{\sqrt{3}}A + \frac{\sqrt{2}}{\sqrt{3}} F
\end{array}
\]
and the product of one $D$ and one $F$ goes like
\[
\begin{array}{ll}
& F \times_{\mathcal{A}_4} D \\
= & \frac{1}{\sqrt{2}} \left(\ket{f}+\ket{\bar{f}}\right)\frac{1}{\sqrt{3}}\left(\ket{d_1}+\ket{d_2}+\ket{d_3}\right) \\
\to & D
\end{array}
\]
This closed algebra at the top boundary constrains the allowed fusion tree of order parameters to ones with branches only in $A$, $D$ and $F$. We will only consider allowed fusion trees in the following discussions.

With $\mathcal{A}_4$ as the top boundary, the horizontal string operators that generate the symmetry of the sandwich are $W_B$ and $W_C$ which satisfy the fusion rule
\[
W_B^2 = W_A, W_C^2 = W_A+W_B+W_C, W_BW_C = W_C
\]
hence giving rise to the $\mathrm{Rep}(S_3)$ symmetry. The order parameters are vertical string operators of $V_D$ and $V_F$. $V_D$ is a three dimensional order parameter, $V_F$ is a two dimensional order parameter. As discussed in section~\ref{sec:non-invertible_Op}, two $D$'s fuse in a symmetric way into identity with a singlet state $\ket{d_1d_1}+\ket{d_2d_2}+\ket{d_3d_3}$. We assign real fields $\phi_{d_k}$, $k=1,2,3$, to the three dimensions $\ket{d_k}$. $F$ is very similar to $C$. Two $F$'s also fuse in a symmetric way into identity with singlet state $\ket{f\bar{f}}+\ket{\bar{f}f}$. We assign a complex field $\psi_f$ to $\ket{f}$ and its complex conjugation $\bar{\psi}_f$ to $\ket{\bar{f}}$. 

The Lagrangian terms come from fusion trees of $D$ and $F$ into identity, which can be obtained by replacing $C$ by $F$ in all the fusion diagrams in section~\ref{sec:non-invertible_Op}. The corresponding quantum state, however, cannot be simply obtained by replacing $\ket{c}$, $\ket{\bar{c}}$ by $\ket{f}$, $\ket{\bar{f}}$, because the duality from $C$ to $F$ involves a nontrivial rotation of the internal dimensions of $D$. We list the quantum states and the resulting Lagrangian terms below explicitly. 

At second order, the singlet states still take the form
\[
\ket{f\bar{f}}+\ket{\bar{f}f}, \ket{d_1d_1}+\ket{d_2d_2}+\ket{d_3d_3}
\]
and the corresponding quadratic terms (non-derivative) are
\[
|\psi_f|^2, \ \phi_{d_1}^2+\phi_{d_2}^2+\phi_{d_3}^2
\]
which is completely analogous to the $\mathcal{A}_3$ case.

At third order, the fusion tree involving three $F$'s is analogous to the one involving three $C$'s in section~\ref{sec:non-invertible_Op}, and results in quantum state
\[
\ket{fff}+\ket{\bar{f}\bar{f}\bar{f}}
\]
and Lagrangian term
\[
\psi_f^3+\bar{\psi}_f^3
\]
The one involving two $D$'s and one $F$ is very different from the one involving two $D$'s and one $C$ in section~\ref{sec:non-invertible_Op}. The corresponding quantum state is
\begin{align*}
&\left(\ket{d_1d_2}+\ket{d_2d_3}+\ket{d_3d_1}\right)\ket{f} \\ &+ \left(\ket{d_2d_1}+\ket{d_3d_2}+\ket{d_1d_3}\right)\ket{\bar{f}}
\end{align*}
It is more convenient to switch to the basis of
\[
\begin{array}{l}
\ket{d_0} = \frac{1}{\sqrt{3}}\left(\ket{d_1}+\ket{d_2}+\ket{d_3}\right) \\
\ket{d_+} = \frac{1}{\sqrt{3}}\left(\ket{d_1}+\omega \ket{d_2} + \bar{\omega}\ket{d_3}\right) \\
\ket{d_-} = \frac{1}{\sqrt{3}}\left(\ket{d_1}+\bar{\omega} \ket{d_2} + \omega \ket{d_3}\right) 
\end{array}
\]
If we label the corresponding fields as $\phi_d$, $\psi_d$, $\bar{\psi}_d$, the second order term of $D$ becomes
\[
\phi_d^2 + 2|\psi_d|^2
\]
and the third order term involving $D$ becomes
\[
\left(\phi_d^2 + (\omega+\bar{\omega}) |\psi_d|^2\right) \left(\psi_f + \bar{\psi}_f\right)
\]
When $F$ is condensed in a certain direction (i.e. $\psi_f$ take on an expectation value), this third order term becomes effectively a quadratic term for $D$ which have different coefficients for $\phi_d$ and $\psi_d$. 

At fourth order, there are three fusion trees. One involves four $F$'s, one involves four $D$'s, and one involves two $D$'s and two $F$'s. The corresponding quantum states are (using the new basis of $D$)
\[
\begin{array}{l}
\ket{d_0d_0d_0d_0}+\ket{d_+d_-d_+d_-}+\ket{d_-d_+d_-d_+} \\
\ket{ff\bar{f}\bar{f}} + \text{permutation}, \\
\left(\ket{d_0d_0}+\ket{d_+d_-}+\ket{d_-d_+}\right)\left(\ket{f\bar{f}}+\ket{\bar{f}f}\right) \\
+ \left(\ket{d_0d_0}+\omega\ket{d_+d_-}+\bar{\omega}\ket{d_-d_+}\right)\ket{\bar{f}\bar{f}} \\
+ \left(\ket{d_0d_0}+\bar{\omega}\ket{d_+d_-}+\omega\ket{d_-d_+}\right)\ket{ff}
\end{array}
\]
leading to Lagrangian terms
\[
\begin{array}{l}
\phi_d^4 + |\psi_d|^4 \\
|\psi_f|^4 \\
\phi_d^2 \left(2|\psi_f|^2 + \psi_f^2 + \bar{\psi}_f^2\right) \\
+ |\psi_d|^2 \left(2|\psi_f|^2 + \left(\omega+\bar{\omega}\right)\left(\psi_f^2+\bar{\psi}_f^2\right)\right)
\end{array}
\]
The last term again becomes effectively a quadratic term for $D$ with different coefficients for $\phi_d$ and $\psi_d$ when $F$ is condensed.

Now let's look at the field theory for the transitions when the bottom boundary of the SymTFT sandwich changes from $\mathcal{A}_2$, $\mathcal{A}_3$, $\mathcal{A}_1$ to $\mathcal{A}_4$.

When the bottom boundary changes from $\mathcal{A}_2 = A+B+2F$ to $\mathcal{A}_4 = A+D+F$, the system transitions from a phase where the $\mathrm{Rep}(S_3)$ symmetry is broken down to $Z_2$ to the fully symmetry breaking phase. $F$ is condensed on both sides of the transition and $D$ is the order parameter driving the transition. Since $F$ is condensed with coefficient 2 in $\mathcal{A}_2$, we set $\psi_f$ to a generic complex value in the Lagrangian. As discussed above some of the mixed third and fourth order terms become effectively quadratic terms of $D$. We will focus on the case where the coefficient of $\phi_d^2$ is smaller than that of $|\psi_d|^2$ and $\phi_d$ is the order parameter driving the transition. The $\psi_d$ field remains gapped and is irrelevant to the transition. Collecting all the relevant terms of $\phi_d$, we arrive at the Lagrangian
\[
\begin{array}{lll}
\mathcal{L} & = & K_{\tau} \left(\partial_{\tau}\phi_{d}\right)^2 + K_x \left(\partial_x \phi_{d}\right)^2 + \\
& & a \phi_{d}^2 + b \phi_{d}^4 + ...
\end{array}
\]
which is exactly the same as $\mathcal{L}_B$ in Eq.~\ref{eq:LB} and describes the Ising transition. 

When the bottom boundary changes from $\mathcal{A}_3 = A+C+D$ to $\mathcal{A}_4 = A+D+F$, the system transitions from a phase where the $\mathrm{Rep}(S_3)$ symmetry is broken down to $Z_3$ to the fully symmetry breaking phase. When the top boundary is $\mathcal{A}_1=A+B+2C$, this corresponds to the nonchiral $Z_3$ transition with a 3-state Potts critical point. Changing the top boundary to $\mathcal{A}_4$ corresponds to gauging the $Z_2$ subgroup of $S_3$ while the dynamics is still described by $\mathcal{L}_C$. Let's see if that's what we get from the generalized GL procedure. 

The $D$ anyon is condensed on both sides of the transition and $F$ is the dynamical order parameter driving the transition. Before the transition, $D$ is condensed in the dimension $\ket{d}_1$. Therefore, we set $\phi_{d_1}$ to be some nonzero constant value and $\phi_{d_2} = \phi_{d_3} = 0$. With this choice, the mixed third order term is zero and the mixed fourth order term becomes effectively the quadratic term of $\psi_f$. Collecting all the dynamical terms of $\psi_f$, we arrive at the field theory
\begin{equation}
\begin{array}{lll}
\mathcal{L} & = &  K_{\tau} |\partial_{\tau}\psi_f|^2 + K_x |\partial_x \psi_f|^2 + \\
& & a |\psi_f|^2 + b (\psi_f^3 + \bar{\psi}_f^3) + c |\psi_f|^4 + ...
\end{array}
\end{equation}
which is exactly the same as $\mathcal{L}_C$ describing nonchiral $Z_3$ transition.

Finally, the transition from $\mathcal{A}_1$ to $\mathcal{A}_4$ involves the condensation of both $F$ and $D$. Both the $\psi_f$ and the $\phi_{d}$ fields are dynamical across the transition and the field theory can contain all the terms listed above. We will not try to analyze it further.

\section{When the prescribed procedure fails}
\label{app:nonfull}

Here we choose the top boundary $\mathcal A_3=A+C+D$ and consider the transition between the bottom boundaries $\mathcal A_1=A+B+2C$ and $\mathcal A_2=A+B+2F$. This transition lies outside the setting considered in the main text (section~\ref{sec:non-invertible_SymTFT}): neither of the two bottom boundaries coincides with the top boundary, and hence neither phase is the fully symmetry-breaking phase of the $\mathrm{Rep}(S_3)$ symmetry.

In the phase corresponding to $\mathcal A_1$, the $W_F$ symmetry is broken while the invertible $Z_2$ subgroup generated by $W_B$ remains unbroken, as discussed in Sec.~\ref{sec:non-invertible_transition}. By contrast, the phase with bottom boundary $\mathcal A_2$ has no nontrivial local order parameters and corresponds to a fully $\mathrm{Rep}(S_3)$-symmetric phase.

This transition is related by generalized gauging to the transition discussed in Sec.~\ref{sec:invertible_transition}, corresponding to changing the top boundary from $\mathcal A_3$ to $\mathcal A_1$. With $\mathcal A_1$ as the top boundary, the transition is from the fully $S_3$ symmetry-breaking phase to the $S_3$ to $Z_3$ symmetry-breaking phase, and is described by the chiral $Z_3$ clock model\cite{Huse1982,Chepiga2019,Maceira2022}. Since generalized gauging should not change the dynamical critical exponent $z$, we expect the gauged transition considered here to remain chiral, with the same $z>1$. Indeed, $z$ can be extracted from the finite-size scaling of the energy gap, $\Delta E\sim L^{-z}$, whose scaling exponent should be preserved under generalized gauging.

However, if we apply the procedure outlined in section~\ref{sec:non-invertible_Op} and \ref{sec:non-invertible_transition} blindly to a SymTFT sandwich with top boundary $\mathcal{A}_3$ and bottom boundary transitioning from $\mathcal{A}_2$ to $\mathcal{A}_1$, we get instead a nonchiral critical theory. The $C$ order parameter is condensed in a generic direction on the $\mathcal A_1$ side of the transition but not in $\mathcal{A}_2$. It therefore drives the transition and we assign it a complex field $\psi_c$. The allowed Lagrangian terms derived in Sec.~\ref{sec:non-invertible_Op} contain all the usual terms of $\psi_c$ but do not include the chiral term $\psi_c \partial_x \bar\psi_c-\bar\psi_c\partial_x\psi_c$ because it is not allowed by the $\mathcal{A}_3$ top boundary. The procedure would therefore predict a nonchiral tetracritical Ising CFT, related by gauging to the three-state Potts CFT. Therefore, we see that the procedure outlined in section~\ref{sec:non-invertible_Op} and \ref{sec:non-invertible_transition} cannot be applied in general for transitions that do not involve a fully symmetry-breaking phase. For this procedure to produce the correct result, it is important that the top boundary matches the bottom boundary after the transition so that the top boundary allows all the fluctuation that drives the transition at the bottom boundary.

\bibliography{references}

@article{cong2016topological,
	author = {Cong, Iris and Cheng, Meng and Wang, Zhenghan},
	journal = {arXiv preprint arXiv:1609.02037},
	title = {Topological quantum computation with gapped boundaries},
	year = {2016}}

@book{Kardar2007,
	author = {Kardar, Mehran},
	place = {Cambridge},
	publisher = {Cambridge University Press},
	title = {Statistical Physics of Fields},
	year = {2007}}

@book{Landau1980,
	address = {Oxford},
	author = {Landau, L. D. and Lifshitz, E. M.},
	edition = {3rd},
	publisher = {Pergamon Press},
	title = {Statistical Physics, Part 1},
	year = {1980}}

@article{Brennen2009,
	author = {Brennen, G K and Aguado, M and Cirac, J I},
	doi = {10.1088/1367-2630/11/5/053009},
	journal = {New Journal of Physics},
	month = {may},
	number = {5},
	pages = {053009},
	title = {Simulations of quantum double models},
	url = {https://doi.org/10.1088/1367-2630/11/5/053009},
	volume = {11},
	year = {2009}}

@article{Beigi2011,
	author = {Beigi, Salman and Shor, Peter W. and Whalen, Daniel},
	date = {2011/09/01},
	doi = {10.1007/s00220-011-1294-x},
	id = {Beigi2011},
	isbn = {1432-0916},
	journal = {Communications in Mathematical Physics},
	number = {3},
	pages = {663--694},
	title = {The Quantum Double Model with Boundary: Condensations and Symmetries},
	url = {https://doi.org/10.1007/s00220-011-1294-x},
	volume = {306},
	year = {2011}}

@article{Kong2020,
	author = {Kong, Liang and Lan, Tian and Wen, Xiao-Gang and Zhang, Zhi-Hao and Zheng, Hao},
	doi = {10.1103/PhysRevResearch.2.043086},
	issue = {4},
	journal = {Phys. Rev. Res.},
	month = {Oct},
	numpages = {53},
	pages = {043086},
	publisher = {American Physical Society},
	title = {Algebraic higher symmetry and categorical symmetry: A holographic and entanglement view of symmetry},
	url = {https://link.aps.org/doi/10.1103/PhysRevResearch.2.043086},
	volume = {2},
	year = {2020}}

@article{Kong2022categories,
	author = {Kong, Liang and Zheng, Hao},
	journal = {Journal of High Energy Physics},
	number = {8},
	pages = {1--44},
	publisher = {Springer},
	title = {Categories of quantum liquids I},
	volume = {2022},
	year = {2022}}

@article{Xu2024,
	author = {Xu, Rongge and Zhang, Zhi-Hao},
	journal = {Physical Review B},
	number = {15},
	pages = {155106},
	publisher = {APS},
	title = {Categorical descriptions of one-dimensional gapped phases with Abelian onsite symmetries},
	volume = {110},
	year = {2024}}

@article{Kong2018,
	author = {Kong, Liang and Zheng, Hao},
	journal = {Nuclear Physics B},
	pages = {140--165},
	publisher = {Elsevier},
	title = {Gapless edges of 2d topological orders and enriched monoidal categories},
	volume = {927},
	year = {2018}}

@article{Lin2023,
	author = {Lin, Ying-Hsuan and Okada, Masaki and Seifnashri, Sahand and Tachikawa, Yuji},
	journal = {Journal of High Energy Physics},
	number = {3},
	pages = {1--43},
	publisher = {Springer},
	title = {Asymptotic density of states in 2d CFTs with non-invertible symmetries},
	volume = {2023},
	year = {2023}}

@misc{Freed2023topological,
	archiveprefix = {arXiv},
	author = {Daniel S. Freed and Gregory W. Moore and Constantin Teleman},
	eprint = {2209.07471},
	primaryclass = {hep-th},
	title = {Topological symmetry in quantum field theory},
	year = {2023}}

@misc{Moradi2022topological,
	archiveprefix = {arXiv},
	author = {Heidar Moradi and Seyed Faroogh Moosavian and Apoorv Tiwari},
	eprint = {2207.10712},
	primaryclass = {cond-mat.str-el},
	title = {Topological Holography: Towards a Unification of Landau and Beyond-Landau Physics},
	year = {2022}}

@article{Chatterjee2023symmetry,
	archiveprefix = {arXiv},
	author = {Chatterjee, Arkya and Wen, Xiao-Gang},
	doi = {10.1103/PhysRevB.107.155136},
	eprint = {2203.03596},
	issue = {15},
	journal = {Phys. Rev. B},
	month = {Apr},
	numpages = {34},
	pages = {155136},
	publisher = {American Physical Society},
	title = {Symmetry as a shadow of topological order and a derivation of topological holographic principle},
	url = {https://link.aps.org/doi/10.1103/PhysRevB.107.155136},
	volume = {107},
	year = {2023}}

@article{Apruzzi2023,
	author = {Apruzzi, Fabio and Bonetti, Federico and Garc{\'\i}a Etxebarria, I{\~n}aki and Hosseini, Saghar S. and Sch{\"a}fer-Nameki, Sakura},
	doi = {10.1007/s00220-023-04737-2},
	issn = {1432-0916},
	journal = {Communications in Mathematical Physics},
	month = may,
	number = {1},
	pages = {895--949},
	publisher = {Springer Science and Business Media LLC},
	title = {Symmetry TFTs from String Theory},
	url = {http://dx.doi.org/10.1007/s00220-023-04737-2},
	volume = {402},
	year = {2023}}

@article{Lichtman2021,
	author = {Lichtman, Tsuf and Thorngren, Ryan and Lindner, Netanel H. and Stern, Ady and Berg, Erez},
	doi = {10.1103/PhysRevB.104.075141},
	issue = {7},
	journal = {Phys. Rev. B},
	month = {Aug},
	numpages = {27},
	pages = {075141},
	publisher = {American Physical Society},
	title = {Bulk anyons as edge symmetries: Boundary phase diagrams of topologically ordered states},
	url = {https://link.aps.org/doi/10.1103/PhysRevB.104.075141},
	volume = {104},
	year = {2021}}

@article{Gaiotto2021,
	author = {Gaiotto, Davide and Kulp, Justin},
	doi = {10.1007/jhep02(2021)132},
	issn = {1029-8479},
	journal = {Journal of High Energy Physics},
	month = feb,
	number = {2},
	publisher = {Springer Science and Business Media LLC},
	title = {Orbifold groupoids},
	url = {http://dx.doi.org/10.1007/JHEP02(2021)132},
	volume = {2021},
	year = {2021}}

@article{Pulmann2021,
	author = {Pulmann, J{\'a}n and {\v S}evera, Pavol and Valach, Fridrich},
	doi = {10.4310/atmp.2021.v25.n1.a5},
	issn = {1095-0753},
	journal = {Advances in Theoretical and Mathematical Physics},
	number = {1},
	pages = {241--274},
	publisher = {International Press of Boston},
	title = {A nonabelian duality for (higher) gauge theories},
	url = {http://dx.doi.org/10.4310/ATMP.2021.v25.n1.a5},
	volume = {25},
	year = {2021}}

@article{Bhardwaj2020,
	author = {Bhardwaj, Lakshya and Lee, Yasunori and Tachikawa, Yuji},
	doi = {10.1007/jhep11(2020)141},
	issn = {1029-8479},
	journal = {Journal of High Energy Physics},
	month = nov,
	number = {11},
	publisher = {Springer Science and Business Media LLC},
	title = {SL(2, Z) action on QFTs with Z2 symmetry and the Brown-Kervaire invariants},
	url = {http://dx.doi.org/10.1007/JHEP11(2020)141},
	volume = {2020},
	year = {2020}}

@article{Bhardwaj2025,
	author = {Lakshya Bhardwaj and Lea E. Bottini and Daniel Pajer and Sakura Sch{\"a}fer-Nameki},
	doi = {10.21468/SciPostPhys.18.1.032},
	journal = {SciPost Phys.},
	pages = {032},
	publisher = {SciPost},
	title = {Gapped phases with non-invertible symmetries: (1+1)d},
	url = {https://scipost.org/10.21468/SciPostPhys.18.1.032},
	volume = {18},
	year = {2025}}

@article{Ji2020,
	author = {Ji, Wenjie and Wen, Xiao-Gang},
	doi = {10.1103/PhysRevResearch.2.033417},
	issue = {3},
	journal = {Phys. Rev. Res.},
	month = {Sep},
	numpages = {24},
	pages = {033417},
	publisher = {American Physical Society},
	title = {Categorical symmetry and noninvertible anomaly in symmetry-breaking and topological phase transitions},
	url = {https://link.aps.org/doi/10.1103/PhysRevResearch.2.033417},
	volume = {2},
	year = {2020}}

@article{Kong2021,
	author = {Kong, Liang and Zheng, Hao},
	journal = {Nuclear Physics B},
	pages = {115384},
	publisher = {Elsevier},
	title = {A mathematical theory of gapless edges of 2d topological orders. Part II},
	volume = {966},
	year = {2021}}

@article{Kong2022one,
	author = {Kong, Liang and Wen, Xiao-Gang and Zheng, Hao},
	journal = {Journal of High Energy Physics},
	number = {3},
	pages = {1--32},
	publisher = {Springer},
	title = {One dimensional gapped quantum phases and enriched fusion categories},
	volume = {2022},
	year = {2022}}

@article{kong2024categories,
	author = {Kong, Liang and Zheng, Hao},
	journal = {Communications in Mathematical Physics},
	number = {9},
	pages = {203},
	publisher = {Springer},
	title = {Categories of quantum liquids II},
	volume = {405},
	year = {2024}}

@article{Kong2020classification,
	author = {Kong, Liang and Lan, Tian and Wen, Xiao-Gang and Zhang, Zhi-Hao and Zheng, Hao},
	doi = {10.1007/jhep09(2020)093},
	issn = {1029-8479},
	journal = {Journal of High Energy Physics},
	month = sep,
	number = {9},
	publisher = {Springer Science and Business Media LLC},
	title = {Classification of topological phases with finite internal symmetries in all dimensions},
	url = {http://dx.doi.org/10.1007/JHEP09(2020)093},
	volume = {2020},
	year = {2020}}

@article{kong2020mathematical,
	author = {Kong, Liang and Zheng, Hao},
	journal = {Journal of High Energy Physics},
	number = {2},
	pages = {1--62},
	publisher = {Springer},
	title = {A mathematical theory of gapless edges of 2d topological orders. Part I},
	volume = {2020},
	year = {2020}}

@misc{Kong2015,
	archiveprefix = {arXiv},
	author = {Kong, Liang and Wen, Xiao-Gang and Zheng, Hao},
	eprint = {1502.01690},
	title = {Boundary-bulk relation for topological orders as the functor mapping higher categories to their centers},
	year = {2015}}

@article{Huse1982,
	author = {Huse, David A. and Fisher, Michael E.},
	doi = {10.1103/PhysRevLett.49.793},
	issue = {11},
	journal = {Phys. Rev. Lett.},
	month = {Sep},
	numpages = {0},
	pages = {793--796},
	publisher = {American Physical Society},
	title = {Domain Walls and the Melting of Commensurate Surface Phases},
	url = {https://link.aps.org/doi/10.1103/PhysRevLett.49.793},
	volume = {49},
	year = {1982}}

@article{Chepiga2019,
	author = {Chepiga, Natalia and Mila, Fr\'ed\'eric},
	doi = {10.1103/PhysRevLett.122.017205},
	issue = {1},
	journal = {Phys. Rev. Lett.},
	month = {Jan},
	numpages = {5},
	pages = {017205},
	publisher = {American Physical Society},
	title = {Floating Phase versus Chiral Transition in a 1D Hard-Boson Model},
	url = {https://link.aps.org/doi/10.1103/PhysRevLett.122.017205},
	volume = {122},
	year = {2019}}

@article{Maceira2022,
	author = {Maceira, Ivo A. and Chepiga, Natalia and Mila, Fr\'ed\'eric},
	doi = {10.1103/PhysRevResearch.4.043102},
	issue = {4},
	journal = {Phys. Rev. Res.},
	month = {Nov},
	numpages = {10},
	pages = {043102},
	publisher = {American Physical Society},
	title = {Conformal and chiral phase transitions in Rydberg chains},
	url = {https://link.aps.org/doi/10.1103/PhysRevResearch.4.043102},
	volume = {4},
	year = {2022}}

@article{Kitaev2003,
	author = {A.Yu. Kitaev},
	doi = {https://doi.org/10.1016/S0003-4916(02)00018-0},
	issn = {0003-4916},
	journal = {Annals of Physics},
	number = {1},
	pages = {2-30},
	title = {Fault-tolerant quantum computation by anyons},
	url = {https://www.sciencedirect.com/science/article/pii/S0003491602000180},
	volume = {303},
	year = {2003}}

@misc{chen2026spontaneous,
	archiveprefix = {arXiv},
	author = {Xie Chen and Shang Liu and Da-chuan Lu and Nathanan Tantivasadakarn},
	eprint = {2605.27672},
	primaryclass = {cond-mat.str-el},
	title = {Spontaneous breaking of non-invertible symmetries and duality to beyond-Landau transitions},
	url = {https://arxiv.org/abs/2605.27672},
	year = {2026}}

@article{Chen2025,
	author = {Chen, Xie},
	doi = {10.1103/tmvy-vsqd},
	issue = {25},
	journal = {Phys. Rev. Lett.},
	month = {Dec},
	numpages = {12},
	pages = {250001},
	publisher = {American Physical Society},
	title = {Essay: Generalized Landau Paradigm for Quantum Phases and Phase Transitions},
	url = {https://link.aps.org/doi/10.1103/tmvy-vsqd},
	volume = {135},
	year = {2025}}

@misc{Lootens2024,
	archiveprefix = {arXiv},
	author = {Laurens Lootens and Clement Delcamp and Frank Verstraete},
	eprint = {2408.06334},
	primaryclass = {quant-ph},
	title = {Entanglement and the density matrix renormalisation group in the generalised Landau paradigm},
	url = {https://arxiv.org/abs/2408.06334},
	year = {2024}}

@article{Kennedy1992a,
	author = {Kennedy, Tom and Tasaki, Hal},
	date = {1992/07/01},
	doi = {10.1007/BF02097239},
	id = {Kennedy1992},
	isbn = {1432-0916},
	journal = {Communications in Mathematical Physics},
	number = {3},
	pages = {431--484},
	title = {Hidden symmetry breaking and the Haldane phase in S=1 quantum spin chains},
	url = {https://doi.org/10.1007/BF02097239},
	volume = {147},
	year = {1992}}

@article{Kennedy1992b,
	author = {Kennedy, Tom and Tasaki, Hal},
	doi = {10.1103/PhysRevB.45.304},
	issue = {1},
	journal = {Phys. Rev. B},
	month = {Jan},
	numpages = {0},
	pages = {304--307},
	publisher = {American Physical Society},
	title = {Hidden ${\mathrm{Z}}_{2}$\ifmmode\times\else\texttimes\fi{}${\mathrm{Z}}_{2}$ symmetry breaking in Haldane-gap antiferromagnets},
	url = {https://link.aps.org/doi/10.1103/PhysRevB.45.304},
	volume = {45},
	year = {1992}}

@misc{Huang2023topological,
	archiveprefix = {arXiv},
	author = {Sheng-Jie Huang and Meng Cheng},
	eprint = {2310.16878},
	primaryclass = {cond-mat.str-el},
	title = {Topological holography, quantum criticality, and boundary states},
	year = {2023}}

@article{Zhang2023,
	author = {Zhang, Carolyn and Levin, Michael},
	doi = {10.1103/PhysRevLett.130.026801},
	issue = {2},
	journal = {Phys. Rev. Lett.},
	month = {Jan},
	numpages = {6},
	pages = {026801},
	publisher = {American Physical Society},
	title = {Exactly Solvable Model for a Deconfined Quantum Critical Point in 1D},
	url = {https://link.aps.org/doi/10.1103/PhysRevLett.130.026801},
	volume = {130},
	year = {2023}}

@misc{Warman2026,
	archiveprefix = {arXiv},
	author = {Alison Warman and Yuhan Gai and Sakura Schafer-Nameki},
	eprint = {2605.31601},
	primaryclass = {cond-mat.str-el},
	title = {Twin Phases: Intrinsic Deconfined Quantum Criticality},
	url = {https://arxiv.org/abs/2605.31601},
	year = {2026}}

@misc{Gai2026,
	archiveprefix = {arXiv},
	author = {Yuhan Gai and Sakura Schafer-Nameki and Alison Warman},
	eprint = {2605.31602},
	primaryclass = {cond-mat.str-el},
	title = {Twin Algebras: Condensable Algebras beyond Anyons},
	url = {https://arxiv.org/abs/2605.31602},
	year = {2026}}

@article{Landau1937,
	author = {Landau, L. D.},
	doi = {10.1016/B978-0-08-010586-4.50034-1},
	editor = {ter Haar, D.},
	journal = {Zh. Eksp. Teor. Fiz.},
	pages = {19--32},
	title = {{On the theory of phase transitions}},
	volume = {7},
	year = {1937}}

@article{Hofman2019,
  author = {Diego M. Hofman and Nabil Iqbal},
  doi = {10.21468/SciPostPhys.6.1.006},
  journal = {SciPost Phys.},
  pages = {006},
  publisher = {SciPost},
  title = {{Goldstone modes and photonization for higher form symmetries}},
  url = {https://scipost.org/10.21468/SciPostPhys.6.1.006},
  volume = {6},
  year = {2019},
}

@article{Delacretaz2020,
  author = {Luca V. Delacr{\'e}taz and Diego M. Hofman and Gr{\'e}goire Mathys},
  doi = {10.21468/SciPostPhys.8.3.047},
  journal = {SciPost Phys.},
  pages = {047},
  publisher = {SciPost},
  title = {{Superfluids as higher-form anomalies}},
  url = {https://scipost.org/10.21468/SciPostPhys.8.3.047},
  volume = {8},
  year = {2020},
}

@article{Iqbal2020,
  author = {Nabil Iqbal and John McGreevy},
  doi = {10.21468/SciPostPhys.9.2.019},
  journal = {SciPost Phys.},
  pages = {019},
  publisher = {SciPost},
  title = {{Toward a 3d Ising model with a weakly-coupled string theory dual}},
  url = {https://scipost.org/10.21468/SciPostPhys.9.2.019},
  volume = {9},
  year = {2020},
}

@article{McGreevy2023,
  author = {McGreevy, John},
  doi = {https://doi.org/10.1146/annurev-conmatphys-040721-021029},
  issn = {1947-5462},
  journal = {Annual Review of Condensed Matter Physics},
  number = {Volume 14, 2023},
  pages = {57-82},
  publisher = {Annual Reviews},
  title = {Generalized Symmetries in Condensed Matter},
  type = {Journal Article},
  url = {https://www.annualreviews.org/content/journals/10.1146/annurev-conmatphys-040721-021029},
  volume = {14},
  year = {2023},
}

@article{Moradi2023,
  title = {Topological holography: Towards a unification of Landau and beyond-Landau physics},
  pages = {066},
  author = {Moradi, Heidar and Moosavian, Seyed Faroogh and Tiwari, Apoorv},
  journal = {SciPost Phys. Core},
  volume = {6},
  year = {2023},
  publisher = {SciPost},
  doi = {10.21468/SciPostPhysCore.6.4.066},
  url = {https://scipost.org/10.21468/SciPostPhysCore.6.4.066},
}

@article{Gaiotto2015,
  author = {Gaiotto, Davide and Kapustin, Anton and Seiberg, Nathan and Willett, Brian},
  date = {2015/02/26},
  doi = {10.1007/JHEP02(2015)172},
  id = {Gaiotto2015},
  isbn = {1029-8479},
  journal = {Journal of High Energy Physics},
  number = {2},
  pages = {172},
  title = {Generalized global symmetries},
  url = {https://doi.org/10.1007/JHEP02(2015)172},
  volume = {2015},
  year = {2015},
}

@misc{Shao2024TASI,
  archiveprefix = {arXiv},
  author = {Shu-Heng Shao},
  eprint = {2308.00747},
  primaryclass = {hep-th},
  title = {What's Done Cannot Be Undone: TASI Lectures on Non-Invertible Symmetries},
  url = {https://arxiv.org/abs/2308.00747},
  year = {2024},
}

@article{SCHAFERNAMEKI2024,
  author = {Sakura Sch{\"a}fer-Nameki},
  doi = {https://doi.org/10.1016/j.physrep.2024.01.007},
  issn = {0370-1573},
  journal = {Physics Reports},
  note = {ICTP lectures on (non-)invertible generalized symmetries},
  pages = {1-55},
  title = {ICTP lectures on (non-)invertible generalized symmetries},
  url = {https://www.sciencedirect.com/science/article/pii/S0370157324000310},
  volume = {1063},
  year = {2024},
}

@article{Bhardwaj2024,
  title = {Categorical Landau Paradigm for Gapped Phases},
  author = {Bhardwaj, Lakshya and Bottini, Lea E. and Pajer, Daniel and Sch\"afer-Nameki, Sakura},
  journal = {Phys. Rev. Lett.},
  volume = {133},
  issue = {16},
  pages = {161601},
  numpages = {6},
  year = {2024},
  month = {Oct},
  publisher = {American Physical Society},
  doi = {10.1103/PhysRevLett.133.161601},
  url = {https://link.aps.org/doi/10.1103/PhysRevLett.133.161601},
}

@article{Li2026,
  title = {Gapped boundaries of Kitaev's quantum double models: A lattice realization of anyon condensation from Lagrangian algebras},
  author = {Li, Mu and Yang, Xiao-Han and Dong, Xiao-Yu},
  journal = {Phys. Rev. B},
  volume = {113},
  issue = {3},
  pages = {035150},
  numpages = {32},
  year = {2026},
  month = {Jan},
  publisher = {American Physical Society},
  doi = {10.1103/lg28-bxyv},
  url = {https://link.aps.org/doi/10.1103/lg28-bxyv},
}

@article{Bhardwaj2026,
  title = {Lattice models for phases and transitions with non-invertible symmetries},
  pages = {134},
  author = {Bhardwaj, Lakshya and Bottini, Lea E. and Schäfer-Nameki, Sakura and Tiwari, Apoorv},
  journal = {SciPost Phys.},
  volume = {20},
  year = {2026},
  publisher = {SciPost},
  doi = {10.21468/SciPostPhys.20.5.134},
  url = {https://scipost.org/10.21468/SciPostPhys.20.5.134},
}

@article{Chatterjee2024,
  title = {Quantum phases and transitions in spin chains with non-invertible symmetries},
  pages = {115},
  author = {Chatterjee, Arkya and Aksoy, {\"O}mer M. and Wen, Xiao-Gang},
  journal = {SciPost Phys.},
  volume = {17},
  year = {2024},
  publisher = {SciPost},
  doi = {10.21468/SciPostPhys.17.4.115},
  url = {https://scipost.org/10.21468/SciPostPhys.17.4.115},
}

@article{Choi2023,
  title = {Remarks on boundaries, anomalies, and noninvertible symmetries},
  author = {Choi, Yichul and Rayhaun, Brandon C. and Sanghavi, Yaman and Shao, Shu-Heng},
  journal = {Phys. Rev. D},
  volume = {108},
  issue = {12},
  pages = {125005},
  numpages = {41},
  year = {2023},
  month = {Dec},
  publisher = {American Physical Society},
  doi = {10.1103/PhysRevD.108.125005},
  url = {https://link.aps.org/doi/10.1103/PhysRevD.108.125005},
}

@article{Pollmann2012a,
  author = {Pollmann, Frank and Turner, Ari M.},
  doi = {10.1103/physrevb.86.125441},
  issn = {1550-235X},
  journal = {Physical Review B},
  month = sep,
  number = {12},
  publisher = {American Physical Society (APS)},
  title = {Detection of symmetry-protected topological phases in one dimension},
  url = {http://dx.doi.org/10.1103/PhysRevB.86.125441},
  volume = {86},
  year = {2012},
}

@article{Kobayashi2026,
  title = {Soft symmetries of topological orders},
  author = {Kobayashi, Ryohei and Barkeshli, Maissam},
  journal = {Phys. Rev. B},
  volume = {113},
  issue = {11},
  pages = {115150},
  numpages = {17},
  year = {2026},
  month = {Mar},
  publisher = {American Physical Society},
  doi = {10.1103/cwn4-jl57},
  url = {https://link.aps.org/doi/10.1103/cwn4-jl57},
}

\end{document}